\documentclass[aps,twocolumn,prb,superscriptaddress]{revtex4-2}

\usepackage{graphicx}
\usepackage{booktabs}
\usepackage{setspace}
\usepackage{amsmath}
\usepackage{amssymb}
\usepackage{amsthm}
\usepackage{tikz}
\usetikzlibrary{shapes.misc}
\usetikzlibrary{positioning}
\usetikzlibrary {arrows.meta}
\usetikzlibrary{calc}
\usepackage{braket}
\usepackage{natbib}
\graphicspath{../2_plots}

\definecolor{rpagreen}{RGB}{0,150,0}

\definecolor{nzorng}{RGB}{255, 35, 250}

\definecolor{willcolor}{RGB}{0, 155, 155}

\begin{document}

\title{Expressivity of determinantal ansatzes for neural network wave functions}

\author{Ni Zhan}
\affiliation{Department of Computer Science; Princeton University; Princeton, NJ 08544}

\author{William A. Wheeler}
\affiliation{Department of Electrical and Biomedical Engineering, University of Vermont; Burlington, VT 05405 
}

\author{Gil Goldshlager}
\affiliation{Department of Mathematics; University of California, Berkeley; Berkeley, CA 94720}

\author{Elif Ertekin}
\affiliation{Department of Mechanical Science and Engineering; The Grainger College of Engineering; University of Illinois Urbana-Champaign; Urbana, Illinois 61801}

\author{Ryan P. Adams}
\affiliation{Department of Computer Science; Princeton University; Princeton, NJ 08544}

\author{Lucas K. Wagner}
\email{lkwagner@illinois.edu}
\affiliation{AJL Institute for Condensed Matter Theory; Department of Physics; The Grainger College of Engineering; University of Illinois Urbana-Champaign; Urbana, Illinois 61801 }

\begin{abstract}
Neural network wave functions have shown promise as a way to achieve high accuracy on the many-body quantum problem. 
These wave functions most commonly use a determinant or sum of determinants to antisymmetrize many-body orbitals which are described by a neural network. 
In many cases, the wave function is projected onto a fixed-spin state.
Such a treatment is allowed for spin-independent operators; however, it cannot be applied to spin-dependent problems, such as Hamiltonians containing spin-orbit interactions.
We show that for spin-independent Hamiltonians, a strict upper bound property is obeyed between a traditional Hartree-Fock like determinant, full spinor wave function, the full determinant wave function, and a generalized spinor wave function. 
The relationship between a spinor wave function and the full determinant arises because the full determinant wave function is the spinor wave function projected onto a fixed-spin, after which antisymmetry is implicitly restored in the spin-independent case.
For spin-dependent Hamiltonians, the full determinant wave function is not applicable, because it is not antisymmetric.
Numerical experiments on the H$_3$ molecule and two-dimensional homogeneous electron gas confirm the bounds.

\end{abstract}
\maketitle

\section{Introduction}

Accurately computing the properties of many-electron systems is a central challenge in quantum chemistry, quantum physics, and materials science \cite{motta2020ground,williams-2020-direc-compar}. \textit{Ab initio} descriptions of strongly correlated matter enable physical insights and understanding of materials exhibiting complex spin textures, superconductivity, superfluidity, and other exotic phases \cite{martin2020electronic,smith2017interacting}. While modeling of many systems assumes spin-independent Hamiltonians, spin-dependent Hamiltonians are important for effects such as magnetism and spin-orbit coupling. High accuracy many-body calculations of spin-dependent Hamiltonians would elucidate new phenomena with applications to materials such as twisted bilayer graphene and transition metal dichalcogenides.

Many-body quantum calculations are challenging, because the Hilbert space of the wave function grows exponentially with increasing system size. The expressivity of the ansatz determines how closely a state in the Hilbert space (e.g., the ground state of the Hamiltonian) can be represented; thus, increasing expressivity achieves more accurate ground states, though at higher computational cost. Recent work has shown that a single determinant with infinitely flexible many-body orbitals completely represents any antisymmetric function~\cite{pfau-2020-initio-solut}. 
Ansatzes using neural networks (NNs) to parameterize the many-body orbitals, trained with variational Monte Carlo, have reached state-of-the-art results because of their high degree of flexibility \cite{hermann-2023-ab-initio,hermann-2020-deep-neural,carleo-2017-solvin-quant,luo-2019-backf-trans,pfau-2020-initio-solut,li-2022-ab-initio,wilson-2023-neural-networ,cassella-2023-discov-quant,smith-2024-unified-variat,yoshioka-2021-solvin-quasip,scherbela-2024-towar-trans,linteau-2025-univer-neural,xie-2023-deep-variat,pescia-2024-messag-passin,scherbela-2025-accur-ab,NEURIPS2022_43089499}. 
In practice, solution convergence is not quick, and a sum of determinants and the ``full determinant" are used to increase variational freedom. 
Many NN wave functions trained in the first quantization project onto a fixed-spin configuration. Despite the relevance of spin-dependent interacting Hamiltonians, there has been less work on these~\cite{melton2016quantum,melton2016spin,schmidt-1999-quant-monte,lonardoni-2018-auxil-field}. The spin contamination of different Slater-Jastrow wave functions was investigated in \citet{huang-1998-spin-contam}. Recently, spin-dependent neural wave functions have been used for ultra-cold Fermi gases~\cite{kim-2024-neural-networ}, nuclear physics~\cite{adams-2021-variat-monte}, and fractional electron fillings in a Moir\'e material~\cite{luo-2025-solvin-fract}, and spin-based penalties in the training loss function have been used in \citet{li-2024-spin-symmet,szabo-2024-improv-penal}.

Early NN wave functions used a generalization of the unrestricted Hartree-Fock ansatz to many-body orbitals, which we call \emph{collinear} in this article \cite{luo-2019-backf-trans,hermann-2020-deep-neural,pfau-2020-initio-solut}. Further work extended this ansatz to the full determinant \cite{pfau-2020-initio-solut,lin-2023-explic-antis} and spin-dependent spinor wave functions~\cite{kim-2024-neural-networ}. Given that the collinear and full determinant are projected onto a fixed-spin and that the orbitals of these ansatzes depend on spin differently, a natural question is the relation and generality between the ansatzes. In particular, the full determinant is widely used in state-of-the-art NN wave functions, and its relation to other quantum chemistry ansatzes has been an open question \cite{schaetzle-2023-deepq}. Our paper provides a theoretical explanation.

We demonstrate that if the Hamiltonian and the many-body orbitals are spin-independent and otherwise identical, then the minimum energy NN ansatzes can be energetically ordered as follows: $E[\text{collinear}] \geq E[\text{spinor}] \geq E[P_S \text{ spinor}] = E[\text{fulldet}]$, where $P_S$ is a projection onto a fixed-spin state. 
If the many-body orbitals are spin-dependent, which we call ``generalized spinor", then we show further that $E[\text{fulldet}] \geq E[\text{gen spinor}]$. 
For spin-dependent Hamiltonians, the fulldet wave function cannot be used since it relies on spin projection. 
We show numerical experiments on model systems H$_3$ and two-dimensional homogeneous electron gas to confirm these results.

\section{Background}
\label{sec:background}
\subsection{Many-body spinor wave function}

The many-body orbital $\phi$ is a function $\mathbb{R}^{3n} \rightarrow \mathbb{C}$ of all $n$ electron positions $R_i := \{ r_i; \{r_{j\setminus i}\}\}$ where $\{r_{j\setminus i}\}$ indicates all electron positions except the $i\text{-th}$ electron (the special particle), and the orbital value is invariant to permutation order of the $\{j \setminus i\}$ electrons. We use $R$ to represent all of the electron positions.

The many-body spin-orbital maps from $\mathbb{R}^{3n} \rightarrow \mathbb{C}^2$
\begin{equation}
\label{eq:sorbital}
\langle R_i | \phi \rangle = 
    \begin{pmatrix}
  \phi_{\uparrow}(R_i) \\
  \phi_{\downarrow}(R_i) \\
\end{pmatrix},
\end{equation}
which may also be written as $\phi_{\uparrow}(R_i)\ket{\uparrow_i} + \phi_\downarrow(R_i)\ket{\downarrow_i}$. The spin-orbital is a two-component function for spin-1/2 particles, as it is the positional wave function tensored into the $S_z$ eigenbasis for spin. Spin-1/2 particles have two $S_z$ eigenstates, commonly called up $\ket{\uparrow}$ and down $\ket{\downarrow}$, and the spin state of an electron, $s_i$, can be represented as a normalized complex two-dimensional vector or superposition of the $S_z$ eigenstates. The Hilbert space of the many-body spin-orbital is $\mathbb{L}^2(\mathbb{R}^{3n}) \otimes \mathbb{C}^2$. 

The spinor determinant \cite{pauli-1927-zur-quant} is 
\begin{equation}
\begin{aligned}
& \Psi_\text{spinor}(R,S) \\ &= 
    \begin{vmatrix}
s_1^\dagger \cdot \begin{pmatrix}
  \phi_{1,\uparrow}(R_1) \\
  \phi_{1,\downarrow}(R_1) \\
\end{pmatrix}
 &   
\ldots
 &
 s_n^\dagger \cdot \begin{pmatrix}
     \phi_{1,\uparrow}(R_n) \\
  \phi_{1,\downarrow}(R_n) \\  
  \end{pmatrix}
  \\ 
  \vdots & \ddots & \vdots \\
  s_1^\dagger \cdot \begin{pmatrix}
  \phi_{n,\uparrow}(R_1) \\
  \phi_{n,\downarrow}(R_1) \\ 
  \end{pmatrix}&
\ldots
  &
   s_n^\dagger \cdot \begin{pmatrix}
    \phi_{n,\uparrow}(R_n) \\
  \phi_{n,\downarrow}(R_n) \\
  \end{pmatrix}
  \\ 
\end{vmatrix}  \\ &= \text{Det}[\braket{ R_i, s_i|\phi_j}] 
\label{eq:spinordet}
\end{aligned}
\end{equation}
where $S$ is all of the electron spins. The spinor determinant is antisymmetric because an exchange of $r_i, s_i$ and $r_j, s_j$ only exchanges columns $i$ and $j$, which results in a minus sign. The many-body orbitals of Eq.~\eqref{eq:spinordet} depend on the positions of all particles but only on the spin of the $i$-th electron. We also describe a generalized version in which the orbitals include a permutation-invariant dependence of all spins in the next section.

\subsection{Ansatzes}
\label{sec:ansatzes}

Commonly used quantum chemistry ansatzes are representable in the spinor form. A collinear ansatz contains spin-orbitals that are fully up or down; i.e., the spin-orbitals are all aligned along the same spin axis. The collinear determinant is 
\begin{equation}\begin{aligned} &\braket{R,S|\Psi_{\textrm{collinear}}}\\
  &=\begin{vmatrix}
s_1^\dagger \cdot \begin{pmatrix}
  \phi_{1, \uparrow}(R_1) \\
  0 \\
\end{pmatrix}
 &   
\ldots
 &
 s_n^\dagger\cdot\begin{pmatrix}
     \phi_{1, \uparrow}(R_n) \\
 0 \\  
  \end{pmatrix}
  \\ 
  \vdots & & \vdots\\
  s_1^\dagger  \cdot\begin{pmatrix}
     \phi_{n_\uparrow,\uparrow}(R_1) \\
 0 \\  
  \end{pmatrix} & \ldots &   s_n^\dagger \cdot\begin{pmatrix}
     \phi_{n_\uparrow,\uparrow}(R_n) \\
 0 \\  
  \end{pmatrix}\\
 s_1^\dagger    \cdot\begin{pmatrix}
     0 \\
 \phi_{1,\downarrow}(R_1) \\  
  \end{pmatrix} & \ldots &  s_n^\dagger \cdot\begin{pmatrix}
     0 \\
  \phi_{1,\downarrow}(R_n) \\  
  \end{pmatrix}\\
    \vdots & & \vdots\\

s_1^\dagger \cdot\begin{pmatrix}
 0 \\
  \phi_{n_\downarrow, \downarrow}(R_1) \\ 
  \end{pmatrix}&
\ldots
  &
s_n^\dagger \cdot \begin{pmatrix}
   0 \\
  \phi_{n_\downarrow, \downarrow}(R_n) \\
  \end{pmatrix}
  \\ 
\end{vmatrix},
\end{aligned}
\end{equation}
where $n_\uparrow$ and $n_\downarrow$ represent the number of up and down orbitals, respectively. Note that the collinear ansatz is the only ansatz discussed here that requires selecting the number of up and down orbitals, which we index by $1, \dots, n_\uparrow$ and $1,\dots, n_\downarrow$. If we evaluate $\braket{\uparrow ... \uparrow \downarrow ... \downarrow|\Psi_\textrm{collinear}}$ with the number of spin-up and spin-down electrons the same as the number of up and down orbitals, the matrix is block diagonal with $\text{Det}[\Psi] = \text{Det}_\uparrow\text{Det}_\downarrow$. For many systems, the lowest energy state is one with spins as half up and half down and the orbitals are chosen as $n_\uparrow = n_\downarrow = n/2$. 

Restricted Hartree-Fock (RHF) is a special case of $\Psi_{\textrm{collinear}}$ with $R_i$ replaced with $r_i$ and $\phi_{i,\uparrow} = \phi_{i,\downarrow}$. The orbitals are single-particle orbitals, and spatial component of the up and down orbitals is the same. Unrestricted Hartree-Fock (UHF) is also a case of this ansatz with $R_i$ replaced with $r_i$. 

A noncollinear ansatz is one in which spin-orbitals are not constrained along a specified spin axis. The spinor determinant of Eq.~\eqref{eq:spinordet} is a noncollinear ansatz, and Generalized Hartree-Fock (GHF) is a subset of spinor determinant with many-body orbitals replaced by single-body orbitals. It is clear that ${\textrm{RHF}~\subset~\textrm{UHF}~\subset~\textrm{GHF}}$, which implies the energy ordering $\textrm{min}_{\Psi \in \text{RHF}} E[\Psi] \ge 
    \textrm{min}_{\Psi \in \text{UHF}} E[\Psi] \ge 
    \textrm{min}_{\Psi \in \text{GHF}} E[\Psi] $ by the variational principle. Similarly, $\textrm{min}_{\Psi \in \text{collinear}} E[\Psi]~\ge~
    \textrm{min}_{\Psi \in \text{spinor}} E[\Psi] $.

The full determinant ansatz was introduced in \cite{pfau-2020-initio-solut,lin-2023-explic-antis} to increase variational freedom. It has been used across several different NN wave functions including \cite{hermann-2020-deep-neural,li-2022-ab-initio} and also referred to as dense determinant by \cite{glehn-2022-self-atten,NEURIPS2022_43089499}. The full determinant ansatz is formed from a precursor wave function as
\begin{equation}
\begin{aligned}
 &\braket{R,S|\Psi_{\textrm{precursor}}}\\&= \begin{vmatrix}s_1^\dagger \cdot
 \begin{pmatrix}
  \phi_{1, \uparrow}(R_1) \\
  0 \\
\end{pmatrix}
 &   
\ldots
 &s_n^\dagger \cdot
 \begin{pmatrix}
    0 \\
  \phi_{1, \downarrow}(R_n) \\  
  \end{pmatrix}
  \\ 
  \vdots & \ddots & \vdots \\
s_1^\dagger \cdot
 \begin{pmatrix}
   \phi_{n, \uparrow}(R_1) \\
0 \\ 
  \end{pmatrix}&
\ldots
  & s_n^\dagger \cdot
\begin{pmatrix}
   0 \\
  \phi_{n, \downarrow}(R_n) \\
  \end{pmatrix}
  \\ 
\end{vmatrix}.
\label{eq:fulldet}
\end{aligned}
\end{equation}
The full determinant ansatz as defined in~\cite{pfau-2020-initio-solut} is then  $|\Psi_{\text{fulldet}}\rangle = \ket{\uparrow ... \uparrow \downarrow ... \downarrow}\braket{\uparrow ... \uparrow \downarrow ... \downarrow|\Psi_\textrm{precursor}}$, which produces a dense matrix of orbitals that is not block diagonal. Note that while collinear and spinor are antisymmetric, full determinant is not. The antisymmetry condition requires that the wave function gain a minus sign under exchange of both position and spin. Consider exchanging $R_1, s_1$ and $R_n, s_n$ in Eq.~\eqref{eq:fulldet}. 
The original columns are of the form 
\begin{equation*}
    s_1^\dagger\cdot\begin{pmatrix}\phi_{j,\uparrow}(R_1)\\ 0 \end{pmatrix} \text{ and } s_n^\dagger\cdot\begin{pmatrix}0\\ \phi_{j,\downarrow}(R_n)\end{pmatrix},
\end{equation*}
while the new columns are of the form 
\begin{equation*}
    s_n^\dagger\cdot\begin{pmatrix}\phi_{j,\uparrow}(R_n)\\ 0 \end{pmatrix} \text{ and } s_1^\dagger\cdot\begin{pmatrix}0\\ \phi_{j,\downarrow}(R_1)\end{pmatrix},
\end{equation*} 
not the same as exchanging the original columns and thereby not guaranteeing antisymmetry. 
However, for spin-independent operators, the expectation value of the antisymmetrized version of full determinant can be evaluated efficiently. 

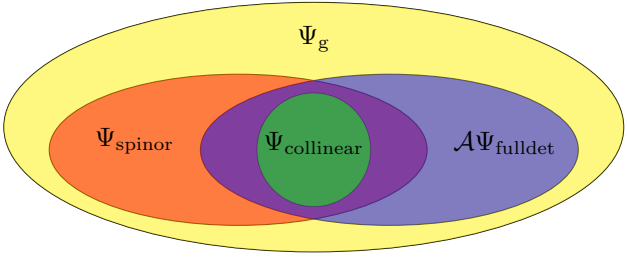
\begin{figure}
\begin{tikzpicture}

\begin{scope}[fill opacity=0.5]
\draw (0.5,0.3) ellipse (4.1cm and 1.65cm);
\draw (-.5,0) ellipse (2.5cm and 1cm);
\draw (1.5,0) ellipse (2.5cm and 1cm);
    \node [draw,
    circle,
    minimum size =1.5cm] at (0.5,0){};

 \fill[yellow] {(0.5,0.3) ellipse (4.1cm and 1.65cm)};
\fill[red] {(-.5,0) ellipse (2.5cm and 1cm)};
\fill[blue] {(1.5,0) ellipse (2.5cm and 1cm)};
    \fill[green] {(0.5,0) circle (0.75cm)};

\end{scope}


    \node [
    label={135:\normalsize$\Psi_\text{spinor}$}] (Al) at (-1.1,-0.3){};


    \node [
    label={45:\normalsize$\mathcal{A} \Psi_\text{fulldet}$}] (Bl) at (2.1,-.3){};

     \node [
    label={\normalsize$\Psi_\text{collinear}$}] (Cl) at (0.5,-0.3){};

    \node [
    label={\normalsize$\Psi_\text{g}$}] (Cl) at (0.5,1.05){};
\end{tikzpicture}
\caption{Containment relations of spinor determinant, antisymmetrized full determinant, collinear, and generalized spinor.}
\label{fig:venn}
\end{figure}

We introduce a generalized spinor which can describe both spinor and full determinant. The generalized spinor has been used in \cite{kim-2024-neural-networ,adams-2021-variat-monte,luo-2025-solvin-fract} and shown here as

\begin{equation}
\begin{aligned}
 &\braket{R,S|\Psi_g}\\&= \begin{vmatrix}s_1^\dagger \cdot
 \begin{pmatrix}
  \phi_{1, \uparrow}(R_1,\{S\}) \\
  \phi_{1, \downarrow}(R_1,\{S\}) \\
\end{pmatrix}
 &   
\ldots
 &s_n^\dagger \cdot
 \begin{pmatrix}
    \phi_{1, \uparrow}(R_n,\{S\}) \\
  \phi_{1, \downarrow}(R_n,\{S\}) \\  
  \end{pmatrix}
  \\ 
  \vdots & \ddots & \vdots \\
s_1^\dagger \cdot
 \begin{pmatrix}
   \phi_{n, \uparrow}(R_1,\{S\}) \\
\phi_{n, \downarrow}(R_1,\{S\}) \\ 
  \end{pmatrix}&
\ldots
  & s_n^\dagger \cdot
\begin{pmatrix}
   \phi_{n, \uparrow}(R_n,\{S\}) \\
  \phi_{n, \downarrow}(R_n,\{S\}) \\
  \end{pmatrix}
  \\ 
\end{vmatrix},
\label{eq:gen-spinor}
\end{aligned}
\end{equation}
where $\{S\}$ indicates a permutation-invariant dependence on $S$ is included in every orbital. When using the overcomplete basis of electron spins as $s_i := \begin{bmatrix} a_i \\b_i \end{bmatrix}, a_i^*a_i^{} + b_i^*b_i^{} = 1$, the wave function must satisfy $\braket{s_i | \Psi} = a_i^* \braket{\uparrow_i | \Psi} + b_i^* \braket{\downarrow_i | \Psi}$ for all $i, R, S$. These constraints of linear superposition are easily satisfied for Eq.~\eqref{eq:gen-spinor} in a discrete and complete basis for spin, e.g.  $\ket\Psi = \frac{1}{\sqrt{2}} \left(\ket{\uparrow\downarrow} + \ket{\downarrow \uparrow}\right)$ for two electrons. However, an arbitrary spin configuration, such as $\braket{s_1s_2|\Psi}$ where $s_i$ are arbitrary spins on the Bloch sphere, requires the evaluation of an exponential number of coefficients, if Eq.~\eqref{eq:gen-spinor} does not satisfy the linearity of spin superposition in the overcomplete basis directly. This results in a physically valid wave function but makes it exponentially expensive to evaluate the wave function at arbitrary superpositions of $S$, prohibiting the usage of the overcomplete basis in practice. This challenge also appeared in \citet{adams-2021-variat-monte,kim-2024-neural-networ}, where they only sampled spins in the discrete basis, which can be inefficient for real space systems.

Fig.~\ref{fig:venn} summarizes the relations between collinear, spinor, full determinant, and generalized spinor. Collinear is fully contained within full determinant, collinear $\subset \textrm{fulldet}$, since collinear must be block diagonal and full determinant need not be. However, the full determinant ansatz is not fully contained within spinor, because the spin-orbitals are not the same across all columns. Spinor is also not fully contained within full determinant since full determinant is always an eigenstate of $S_z$, while the spinor may or may not be. Therefore, the ordering of variational minima between spinor and full determinant is nonobvious. Generalized spinor contains spinor since choosing the orbitals that are independent of spin recovers the spinor determinant form. We will show that generalized spinor also contains full determinant in Sec.~\ref{sec:gen-spinor-pf} and further discuss the validity of non-antisymmetric full determinant and energetic bounds in the remainder of the paper.

\subsection{Spin projection}
For a state in the spin and position basis of $\mathbb{L}^2(\mathbb{R}^{3n})~\otimes~\mathbb{C}^{2n}$, we use the notation
\begin{equation}
    \ket{R, S} = \ket{R}\ket{S}
\end{equation}
so that
\begin{align}
    \braket{S|R, S} &= \ket{R}
 \\ \braket{R|R, S} &= \ket{S}.
\end{align}
 
The spin projection operator $P_S = \ket{S}\bra{S}$ projects a wave function onto the spin state $\ket{S}$. 
The result of a spin projection is a fixed-spin state, where each electron is projected to $\uparrow$ or $\downarrow$, e.g., $\ket{\uparrow \downarrow}\braket{\uparrow \downarrow|\Psi}$. Another example of a fixed-spin state is the spinor wave function 
\begin{equation}
\begin{vmatrix}  \phi_{1,\uparrow}(R_1) & \phi_{1, \downarrow}(R_2)\\ \phi_{2,\uparrow}(R_1) & \phi_{2, \downarrow}(R_2) \end{vmatrix} \ket{\uparrow \downarrow}. 
\end{equation}
A spin-projected wave function is not fully antisymmetric because it does not allow exchange of electrons with different spins. The full determinant as introduced in \cite{pfau-2020-initio-solut} is written for a fixed-spin state $\ket{\uparrow ... \uparrow \downarrow ... \downarrow}$ with $n_\uparrow$ up electrons and $n_\downarrow$ down electrons. 

The correct resolution of the identity for computing expectation values is $\int dR \sum_S \ket{R,S}\bra{R,S}$, which requires summing over all $2^n$ possible spin states. The sum over spin states is commonly omitted by projecting onto a fixed-spin state, which is valid when the antisymmetrized projected wave function has the same expectation as the spin-projected wave function. We show the exact statement and proof in the next section.

\subsection{Evaluating expectation values of the Hamiltonian}

The expectation value of a general operator $O$ (that may be spin-dependent and nonlocal) is

\begin{equation}
\begin{aligned}
\braket{\Psi|O|\Psi} &= \int d R d R^{\prime} 
\\
&\sum_{S, S^{\prime}} \braket{\Psi | R, S} \braket{R, S | O | R^{\prime}, S^{\prime}} \braket{R^{\prime},S^{\prime}|\Psi} 
\label{eq:full-expectation}
\end{aligned}
\end{equation}
where $S$ and $S^{\prime}$ run over all configurations of the spins of all the electrons. 

If $O$ is independent of spin, it has the simplified expectation value
\begin{equation}
    \braket{R, S | O | R^{\prime}, S^{\prime}} 
    = 
    \delta_{SS^{\prime}} \braket{R | O | R^{\prime}},
\label{eq:spin-ind-delta-simplify}
\end{equation}
removing the need to sum over secondary spin configurations. Applying this property to the expectation value in Eq.~\eqref{eq:full-expectation},
\begin{equation}
\begin{aligned}
    \braket{\Psi|O|\Psi} 
    &= \int d R d R^{\prime}  \sum_{S} \braket{\Psi | R, S} \braket{R | O | R^{\prime}} \braket{R^{\prime}, S | \Psi}. 
    \label{eq:spin-independent-Hamiltonian2}
\end{aligned}
\end{equation}

The following theorem states that the spin-projected wave function has the same expectation as its antisymmetrized projected version for a spin-independent operator \cite{foulkes-2001-quant-monte}. 

\newtheorem{theorem}{Theorem}
\begin{theorem}
\label{thm-fix-spin}
If $\Psi$ can be written as an antisymmetrized fixed-spin wave function $\Psi = \mathcal{A}\Psi_{S}$ for some spin configuration $S$, then the expectation can be evaluated just from the projection onto $S $,
\begin{equation}
    \braket{\Psi|O|\Psi} = \int d R d R^{\prime} \braket{\Psi_{S} | R}\braket{R | O | R^{\prime}}   \braket{R^{\prime} | \Psi_{S}}.
\end{equation}
\end{theorem}

\begin{proof}

We summarize the proof which has been shown in Sec.~IV.E of \cite{foulkes-2001-quant-monte}. Consider the fixed-spin wave function $\Psi_{S}$ having spin state $S$.
Its antisymmetrized counterpart is a sum over permutations
\begin{equation}
    \mathcal{A}\Psi_{S} = \frac{1}{n!}\sum_\pi (-1)^\pi P_\pi \Psi_{ S}.
    \label{eq:antisymmetrizer}
\end{equation}
The permutation $P_\pi$ results in a new spin state $S^\pi$.
Since $\braket{S|S^\pi}~=~0$, each permutation $\pi$ contributes a separate integral to the expectation value,
\begin{equation}
\begin{aligned}
    \braket{\mathcal{A}\Psi_{S} | O | \mathcal{A}\Psi_{S}}
    &= \frac{1}{n!}\sum_\pi 
    \int dR dR^\prime \\&\braket{\Psi_{ S^\pi} | R} \braket{R|O|R^\prime}\braket{R^\prime | \Psi_{ S^\pi}}.
    \end{aligned}
\end{equation}
Note that the $(-1)^\pi$ sign terms cancel out, so all contributions are positive.
Since permuting integration variables changes nothing and $\Psi_{S^\pi}$ is the same for all permutations, each term contributes
    $\braket{\Psi_{S} | O | \Psi_{S}}$
up to a normalization factor, equal to the expectation value of the original $\Psi_{S}$.

\end{proof}

The Hamiltonian is a semilocal operator (containing the differential operator for kinetic energy), 
meaning $\braket{R|H\Psi}$ can be evaluated without the extra integral over $R^{\prime}$. Combining this with spin independence of Eq.~\eqref{eq:spin-independent-Hamiltonian2} yields the standard formulation
\begin{equation}
    \braket{\Psi|H|\Psi} = \int d R \braket{\Psi|R}\braket{R | H\Psi},
\end{equation}
which is independent of the spin details of $\Psi$ (assuming a fixed total spin eigenstate). Hence, spin-projection is common and convenient for spin-independent Hamiltonians but generally invalid for spin-dependent Hamiltonians.

\section{Analysis}
\label{sec:spin-project-spinor}
We will show that the full determinant wave function is equivalent to the projection of a spinor determinant onto a particular spin state.
The spin-projected determinant gives the correct expectation value for a \emph{spin-independent} Hamiltonian, but not for general \emph{spin-dependent} Hamiltonians. We show this for two electrons and single-particle orbitals and the general case with many-electrons and many-body orbitals. We also establish that the full determinant (projected spinor) is a variational lower bound to the spinor for spin-independent Hamiltonians.  

 \subsection{Two electrons with single-particle orbitals}
\label{sec:two-elec}
As a concrete example, we first consider a two-electron wave function composed of single-particle orbitals.
We consider the general case of many-electrons and many-body orbitals in the next subsection.

The full determinant  wave function for two electrons and single-particle orbitals is 
\begin{equation}
\ket{\Psi_{\text{fulldet}}} = 
 \begin{vmatrix}  \phi_{1,\uparrow}(r_1) & \phi_{1, \downarrow}(r_2)\\ \phi_{2,\uparrow}(r_1) & \phi_{2, \downarrow}(r_2) \end{vmatrix} \ket{\uparrow \downarrow}.
\label{eq:fulldet-2elec}
\end{equation}

As mentioned in Sec. \ref{sec:ansatzes}, this wave function is not fully antisymmetric because the orbital functions are different across the columns. We also note this wave function is a projection of the spinor determinant onto the  $\ket{\uparrow \downarrow} $ spin state. Using the antisymmetrizer $\mathcal{A}~=~\frac{1}{n!}~\sum_\pi(-1)^\pi~P_\pi$, where $\pi$ represents permutations and $P_\pi$ represents permutation operator (permuting both positions and spins), the full determinant is antisymmetrized to 

\begin{equation}
\begin{aligned}
    &\mathcal{A}\Psi_{\rm fulldet} =  \Psi_{\text{fulldet}, \uparrow \downarrow} - \Psi_{\text{fulldet}, \downarrow \uparrow}\\&=
 \begin{vmatrix}

  \phi_{1, \uparrow}(r_1) 

 &

  \phi_{1, \downarrow}(r_2)
  \\

   \phi_{2, \uparrow}(r_1)
  &

  \phi_{2, \downarrow}(r_2) 
  \\ 
\end{vmatrix} \ket{\uparrow \downarrow} -   \begin{vmatrix}

  \phi_{1, \uparrow}(r_2)
 &

  \phi_{1, \downarrow}(r_1)
  \\

   \phi_{2, \uparrow}(r_2) 

  &

  \phi_{2, \downarrow}(r_1) 
  \\ 
\end{vmatrix} \ket{\downarrow \uparrow}
\label{eq:antifulldet-2elec}
\end{aligned}
\end{equation}
up to normalization.

Now we will show $\braket{\mathcal{A}\Psi_{\textrm{fulldet}}|H|\mathcal{A}\Psi_{\textrm{fulldet}}} = \braket{\Psi_{\textrm{fulldet}}|H|\Psi_{\textrm{fulldet}}}$. Assuming that $H$ is spin-independent (Eq.~\eqref{eq:spin-ind-delta-simplify}) and semilocal, the expectation of Eq.~\eqref{eq:full-expectation} can be simplified to 
\begin{equation}
\begin{aligned}
  &\int dR \sum_S \braket{\Psi | R, S}\braket{R,S|H\Psi}.\\
    \end{aligned}
    \label{eq:sum-s-h-spin-ind}
\end{equation}

For $\mathcal{A}\Psi_{\textrm{fulldet}}$, Eq.~\eqref{eq:sum-s-h-spin-ind} simplifies to 

\begin{equation}
\begin{aligned}
 &\frac{1}{2}  \int dr_1 dr_2 \Psi^*_{\text{fulldet}, \uparrow\downarrow}(r_1, r_2)  [H\Psi_{\text{fulldet}, \uparrow\downarrow}](r_1, r_2)
    \\&\qquad +\frac{1}{2}  \int dr_1 dr_2 \Psi^*_{\text{fulldet}, \downarrow\uparrow}(r_2, r_1) [H \Psi_{\text{fulldet}, \downarrow\uparrow}](r_2, r_1).
    \end{aligned}
    \label{eq:expand-antisymmfull-det-exp}
\end{equation}
The first term is the same as the original full determinant (Eq.~\eqref{eq:fulldet-2elec}), and the second term is equivalent by exchanging the integration variables $r_1$ and $r_2$. This shows that $\Psi_{\textrm{fulldet}}$ has the same energy expectation as    $\mathcal{A}\Psi_{\textrm{fulldet}}$.

Eq.~\eqref{eq:antifulldet-2elec} is a multideterminant wave function. Since it is the antisymmetrized projected spinor determinant and has the same expectation as the full determinant for spin-independent operators, the full determinant has multideterminant character. In the next two sections, we will show that this finding holds for many-electrons and many-body orbitals and that the antisymmetrized projected spinor wave function has a lower minimum energy expectation compared to the spinor wave function.

\subsection{Full determinant is spin-projected spinor}
\label{sec:many-elec-fulldet-is-projected}

Now we show that $\ket{\Psi_{\text{fulldet, S}}}\propto P_S \ket{\Psi_{\textrm{spinor}}}$. 
For each determinant entry,
\begin{equation}
    s_i \cdot \phi_j(R_i) = 
    \phi_{j, \uparrow}(R_i) \braket{s_i | \uparrow_j}
    +\phi_{j, \downarrow}(R_i) \braket{s_i | \downarrow_j}.
\end{equation}
Clearly only the term matching the spin $s_i$ is nonzero, resulting in the determinant
\begin{equation}
\begin{aligned}
    \braket{R, S | \Psi_{\textrm{spinor}}} 
    &=
    \begin{vmatrix}
        \phi_{1,\uparrow}(R_1) 
        & \ldots 
        &    \phi_{1,\downarrow}(R_{n_\uparrow+1})
        & \ldots \\
        \ldots & \ldots & \ldots & \ldots \\
          \phi_{n,\uparrow}(R_1)
        & \ldots 
        &  \phi_{n,\downarrow}(R_{n_\uparrow+1}) & \ldots \\
    \end{vmatrix}\\ &= \braket{ R, S | \Psi_{\textrm{fulldet},S}}.
\end{aligned}
\end{equation}
This projected wave function may no longer be normalized. For projection onto a fixed-spin configuration $S$, the normalization factor is
\begin{equation}
w_S = \frac
{\int dR |\braket{R, S| \Psi_{\textrm{spinor}}} |^2}
{\sum_{S'} \int dR |\braket{R, S^{\prime}| \Psi_{\textrm{spinor}}}  |^2}, 
\end{equation}
making the normalized wave function
\begin{equation}\ket{\Psi_{\textrm{fulldet},S}}
 = \frac{1}{\sqrt{w_S}}  P_S\ket{\Psi_{\textrm{spinor}} }. 
 \label{eq:fulldet-projection}
\end{equation}
Note that $w_S$ is always positive and $\sum_S w_S = 1$. 

We have shown that the full determinant is a spin-projected spinor. By Theorem \ref{thm-fix-spin}, the full determinant has the same expectation as its antisymmetrized version for a spin-independent operator. The antisymmetrized full determinant 
is not the same as the spinor determinant. From the definition of the antisymmetrizer $\mathcal{A}$, the antisymmetrized spin-projected spinor $\mathcal{A}\ket{\Psi_{\text{fulldet}, S}}$ is a multideterminant wave function with up to $n!$ determinants, in contrast to the single determinant $\ket{\Psi_{\text{spinor}}}$.

\subsection{Full determinant is lower bound to spinor determinant energy for spin-independent Hamiltonians}
\label{sec:bounds}

We show that the expectation value of spinor determinant energy is an upper bound to the full determinant energy expectation for a spin-independent Hamiltonian.

\begin{theorem}
\label{thm-bound}
If $H$ operates as the identity in the spin space (i.e. $H$ is spin-independent), then 

\begin{equation}
\braket{\Psi_\text{\rm spinor}|H|\Psi_\text{\rm spinor}}\geq \braket{\Psi_\text{\rm fulldet}|H|\Psi_\text{\rm fulldet}}. 
\end{equation}
\end{theorem}

\begin{proof}
\begin{equation}
\begin{aligned}\braket{\Psi_\text{spinor}|H|\Psi_\text{spinor}}
     &= \sum_{S, S'} \braket{\Psi_{\text{spinor}} | S} \braket{S | H | S'} \braket{S' | \Psi_{\text{spinor}}} \\
&= \sum_{S} \braket{\Psi_{\text{spinor}} | S} H_S \braket{S | \Psi_{\text{spinor}}} \\
&=\sum_{S} w_S \braket{\Psi_{\textrm{fulldet},S}  | H_S|  \Psi_{\textrm{fulldet},S}} \\
&\ge \min_S \braket{\Psi_{\textrm{fulldet},S}  | H_S|  \Psi_{\textrm{fulldet},S}},
\end{aligned}
\end{equation}
where we have used $\braket{S|H|S'} = \delta_{S,S'}H_S$ for spin-independent $H$, Eq.~\eqref{eq:fulldet-projection}, and $w_S\geq 0$.
\end{proof}

We emphasize that the bounds apply across the ansatzes for a fixed orbital expressivity.

\subsection{Generalized spinor contains spinor and full determinant}
\label{sec:gen-spinor-pf}

It is clear that the generalized spinor recovers the spinor by simply dropping the extra dependence on $\{S\}$ from the orbitals. The generalized spinor also generalizes the antisymmetrized full determinant:

\begin{theorem}
Consider the generalized spinor determinant $\Psi_g$ with orbitals
\label{thm-generalized}
\begin{equation}
\phi_{j,\alpha}(R_i,\{S\}) = \phi_{j,\alpha}(R_i) \mathbf{I}_{n_\uparrow,n_\downarrow} (S), \label{eq:select}
\end{equation}
where $\alpha\in \{\uparrow,\downarrow\}$ and $\mathbf{I}_{n_\uparrow,n_\downarrow} (S)$ is the indicator variable for the event that $S$ contains exactly $n_\uparrow$ up-spins and $n_\downarrow$ down-spins. Then $\Psi_g$ is equivalent to an antisymmetrized full determinant with orbitals $\phi_{j,\alpha}(R_i).$
\end{theorem}

For the purpose of representing $\Psi_\textrm{fulldet}$, we only need to consider collinear spins to match $\phi_{j, \alpha}(R_i)$ and $\phi_{j, \alpha}(R_i, \{S\})$.
In addition to clarifying the relation between $\Psi_g$ and $\Psi_\text{fulldet}$, this theorem indicates the following on the full determinant ansatz itself. First, the antisymmetrized full determinant can be represented as a single determinant wave function. Second, in the antisymmetrized full determinant, the spin-projection and  antisymmetrization appear in the form of an indicator function for the desired spin configuration. This observation is an additional viewpoint to the fact that the full determinant is a spin-projected spinor.

\begin{proof}
Let $\Psi_g$ be the generalized spinor determinant with orbitals given by Eq.~(\ref{eq:select}) and $\Psi_f$ be the full determinant with orbitals $\phi_{j,\alpha}(R_i)$. Additionally, let $S_f=\ket{\uparrow ... \uparrow \downarrow ... \downarrow}$ be the spin state with $n_\uparrow$ up-spins followed by $n_\downarrow$ down-spins. We now verify the equivalence $\Psi_g = \mathcal{A} \Psi_f$ by checking all collinear spin-states $S$ using three cases.

Case 1: when $S=S_f$ it holds
\begin{equation}
\braket{R,S_f|\Psi_g} = \braket{R,S_f|\Psi_f} = \braket{R,S_f | \mathcal{A} \Psi_f},
\end{equation}
with the first equality holding since $\mathbf{I}_{n_\uparrow,n_\downarrow} (S_f)=1$ and the second equality holding since $\braket{PR, PS_f | \Psi_f}=0$ when $P$ is a permutation other than the identity.

Case 2: when $S = PS_f$ for some permutation $P$, the equivalence holds by permuting the input state, utilizing the previous case, and permuting it back:
\begin{equation}
\begin{aligned}
\braket{R, PS_f | \Psi_g} &= \mathrm{sgn}(P) \braket{P^{-1}R, S_f | \Psi_g} \\
&= \mathrm{sgn}(P) \braket{P^{-1} R, S_f | \mathcal{A} \Psi_f} \\
&= \braket{R, PS_f | \mathcal{A} \Psi_f}.
\end{aligned}
\end{equation}

Case 3: when $S \neq PS_f$ for any permutation $P$, it holds
\begin{equation}
\braket{R,S | \Psi_g} = \braket{R,S | \mathcal{A} \Psi_f} = 0.
\end{equation}
\end{proof}

We showed that the full determinant is both a projection of the spinor (in Sec.~\ref{sec:many-elec-fulldet-is-projected}), and a specific instance of the generalized spinor. In many NN ansatzes, the many-body spin-orbital depends on the spin of all electrons in a permutationally invariant way, as described by generalized spinor. However, with a fixed-spin projection, the permutation invariant dependence becomes equivalent to a dependence on only the spin of the $i$-th electron, as in the spinor. 

\section{Numerical Experiments}
\label{sec:numerical-exp}

We implement the spinor, projected spinor, and collinear ansatzes to test the energetic bounds for $\text{H}_3$ molecule and 2D homogeneous electron gas. The spinor and projected spinor ansatzes have nonzero spin-orbital components for up and down orbitals while the collinear ansatz has fully up or down spin-orbitals that are spatially unrestricted. 
For the spinor ansatz, we sample each electron's spin, as a normalized complex two-vector, in the Markov chain. 
Hence, the electrons are allowed to have any spin on the Bloch sphere, and we propose spin moves from a von Mises-Fisher proposal distribution \cite{fisher-1953-dispersion} centered at each electron's current spin on the sphere. Discrete sampling of spin would suffice for the integration, but as an overcomplete representation, sampling spin on the Bloch sphere has the same expectation value and often results in faster mixing \cite{melton2016quantum,melton2016spin}. 
In the Markov chain, we alternate spin and position moves while keeping the other fixed. 
The generalized spinor is expensive to evaluate using continuous sampling, and since discrete sampling of spin is inefficient, we leave testing of that result to future work.
Details of sampling and optimization are provided in the Supplemental Material. Data to reproduce the numerical results are provided in \cite{zhan_2025_data}.

\subsection{Bounds on ansatz energy for an $\text{H}_3$ molecule}

\begin{figure}[ht]
\includegraphics{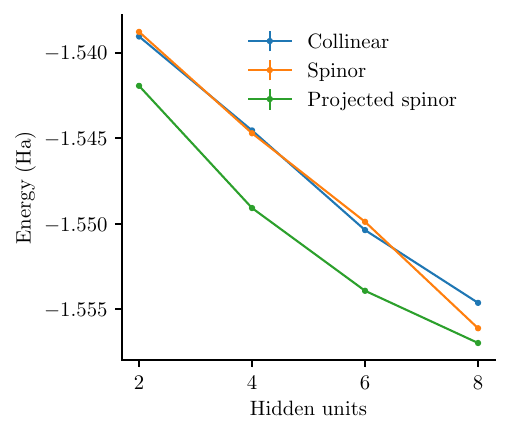}
\caption{Numerical experiments showing the relationship between wave function ansatz on an equilateral
triangle molecule $\text{H}_3$, with a bond length of 2.5 Bohr. At each given orbital expressivity, the energy
relation $E[\text{collinear}]\geq E[\text{spinor}]\geq E[P_S\text{ spinor}]$ is satisfied for all optimized wave functions, and energy difference decreases as orbital expressivity increases.}
\label{fig:h3-bounds}
\end{figure}

We provide numerical evidence of the bound $E[\text{collinear}]\geq E[\text{spinor}]\geq E[P_S\text{ spinor}]$ for $\text{H}_3$, and show that the energy differences decrease with increasing NN capacity. Fig.~\ref{fig:h3-bounds} shows the minimum energy vs. orbital expressivity for $\text{H}_3$ and the spinor, projected spinor, and collinear ansatzes. For this experiment, we simulated $\text{H}_3$ with open boundary conditions and used the FermiNet architecture with single stream features only and two hidden layers with 2, 4, 6, and 8 hidden units per layer. We used these small NNs because energy differences between the three ansatzes diminish with increasing orbital expressivity and are unresolvable when double stream features are included for the $\text{H}_3$ system. In machine learning, it is well known that small neural networks are prone to high variance in their local minima across independent optimizations \cite{frankle-2018-lotter-ticket-hypot,lawrence1997lessons,NEURIPS2019_6a61d423,pmlr-v97-allen-zhu19a,pmlr-v97-du19c}. To avoid bias from local minima, we trained five independent random seeds for each ansatz and ran a separate MCMC evaluation on the minimum energy wave function across the seeds for the reported energy. The batch size was 32,000, and optimizations were run for 40,000 iterations. Fig.~\ref{fig:h3-bounds} shows that the projected spinor is a lower bound in energy to spinor and collinear, and that spinor is a lower bound to collinear, given a fixed orbital expressivity. Although in our numerical experiment, the collinear, spinor, and projected spinor seem to converge in energies, this could be specific to our case, and collinear could be higher energy than spinor in other
systems \cite{jimenez-hoyos-2014-polyr-charac}. Prior work also found that the collinear remained higher in energy than projected spinor for single determinants and certain systems, even in the large NN regime \cite{lin-2023-explic-antis}.

\subsection{Spin-independent and corresponding spin-dependent system}

\begin{figure}[ht!]
\hspace*{-0.4cm}    
\includegraphics{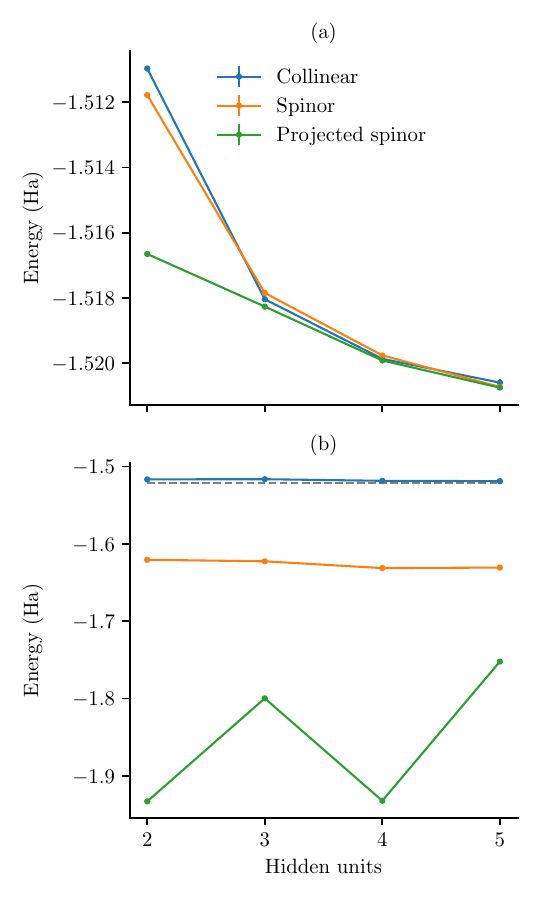}
\caption{Numerical experiments on the periodic 2D homogeneous electron gas, with $r_s=5$, for (a) spin-independent Hamiltonian and (b) spin-dependent Hamiltonian with a Rashba term, $\lambda=0.1$. (a): At each given orbital expressivity, the energy
relation $E[\text{collinear}]\geq E[\text{spinor}]\geq E[P_S\text{ spinor}]$ is satisfied in the spin-independent case. (b): In the spin-dependent case, collinear is unable to model spin effects, spinor correctly models spin effects, and projected spinor is non-variational. Gray dashed line on spin-dependent plot indicates lowest energy from spin-independent plot.}
\label{fig:heg-bounds}
\end{figure}

We show the energetic bounds for a larger spin-independent system and the failure of spin projection for a corresponding spin-dependent system, using the 2D homogeneous electron gas (2DEG) with and without Rashba interaction. The Hamiltonian for the system is 
\begin{equation}
H = \sum_{i=1}^n \frac{P_i^2}{2m} + \lambda \sum_{i=1}^n (p_i^y \sigma_i^x - p_i^x \sigma_i^y) + V_{Coul}(R),
\end{equation}
where $\lambda$ determines the strength of the spin-orbit coupling (Rashba) term. We simulate 10 electrons at $r_s=5$, setting $\lambda=0$ for the spin-independent case and $\lambda=0.1$ for the spin-dependent case. We use the FermiNet architecture with single and double stream features, each with two hidden layers and 2, 3, 4, and 5 hidden units per layer. For the spin-independent case, we trained five independent random seeds for each ansatz and started two trainings of spinor and projected spinor from trained weights of a collinear optimization. For the spin-dependent case, we trained two independent random seeds for each ansatz. The reported energy is from the inference run of the minimum energy wave function across the runs, since we are interested in the lowest energy for the ansatzes. Although the spinor and projected spinor can achieve lower energies than the collinear in principle, a lower energy wave function is not necessarily easier to find during optimization, especially for this particular system, setup, and the small NNs used. The batch size was 2,048 and optimizations were run for 90,000 iterations.

Fig.~\ref{fig:heg-bounds}a shows the energy vs.\ orbital expressivity for the collinear, spinor, and projected spinor ansatzes. The spin-independent case confirms the expected bound, and shows that the bound becomes equality for three or more hidden units. For two hidden units, we expect optimizing more seeds would achieve lower energies for collinear and spinor. Including double stream features in the network shows that the bound also holds for many-body orbitals and the increase in expressivity decreases the difference in bound more rapidly. For a spin-independent Hamiltonian, it is more convenient to use a spin-projected ansatz to decrease computation from spin sampling.

Fig.~\ref{fig:heg-bounds}b shows that spin projection is incorrect for spin-dependent Hamiltonians. When the Rashba term is included, the collinear ansatz is unable to capture Rashba energy, and the projected spinor achieves a non-variational energy. Both the collinear and projected spinor wave functions actually evaluate the wrong Hamiltonian: their expectation value is $\braket{\Psi|H_S|\Psi}$, where $H_S$ is a partial projection of $H$ onto an $S$ subspace, whereas the evaluation for the spinor is  $\braket{\Psi|H|\Psi}$, the correct expectation of the full Hamiltonian. The energy per electron achieved by the spinor ansatz ($-0.163$ Ha) is close to a reference of $-0.15775$ Ha for the same system with 58 electrons in \cite{ambrosetti-2009-quant-monte}. For spin-dependent Hamiltonians, the spinor determinant is a general ansatz that is variational and captures spin effects. 

Demonstrating the bounds on a spin-independent system is a useful test to check the optimization and implementation. Best practices for optimizing NN wave functions is an active area of research \cite{goldshlager-2024-kaczm-inspir}, and NN wave functions have been found to achieve varying energies depending on the optimization settings \cite{glehn-2022-self-atten}. In extending wave functions to spin-dependent systems, since the bound only applies to spin-independent systems, achieving the bound on a corresponding spin-independent system can be a useful first step and provide some information about implementation correctness. 

\section{Conclusions}
\label{sec:conclusions}

We established a hierarchy of expressivity among the collinear, spinor, full determinant, and generalized spinor ansatzes. For spin-independent Hamiltonians and  many-body orbitals that do not depend on spin, the minimum energies are ordered as: $E[\text{collinear}] \geq E[\text{spinor}] \geq E[P_S \text{ spinor}] = E[\text{fulldet}]$. If the many-body orbitals do depend on spin, then we obtain $E[\text{fulldet}] \geq E[\text{gen spinor}]$. We showed these relationships arise because the full determinant is a projection of spinor onto a fixed-spin state and also a specific instance of the generalized spinor. These results provide a new theoretical explanation for the commonly observed fact that full determinants are more expressive than collinear determinants in practice. We implemented the spinor NN wave function and confirmed the bounds for spin-independent Hamiltonians. Additionally, the spinor correctly captured spin-dependent energy, while full determinant is invalid since it relies on spin-projection. Our work opens the path to realistic simulation of both electron correlation and spin-dependent properties in materials.

\vspace{0.6cm}

\section{Acknowledgements}
\label{sec:acknowledgements}

This work was partially supported by NSF OAC 2118201. N.Z. acknowledges support from the Princeton $\text{AI}^2$ initiative. G.G. acknowledges support from the U.S. Department of Energy, Office of Science, Office of Advanced Scientific Computing Research, Department of Energy Computational Science Graduate Fellowship under Award Number DE-SC0023112.
The contributions of L.K.W. in supervising, writing, and creating the theory were supported by the U.S. Department of Energy, Office of Science, Office of Basic Energy Sciences, Computational Materials Sciences Program, under Award No. DE-SC0020177. The authors thank anonymous reviewers for their helpful comments and suggestions. This work used Princeton neuronic cluster and Delta GPU at the National Center for Supercomputing Applications through allocation MAT220011 from the Advanced Cyberinfrastructure Coordination Ecosystem: Services \& Support (ACCESS) program, which is supported by National Science Foundation grants \#2138259, \#2138286, \#2138307, \#2137603, and \#2138296.

\section*{Disclaimer}
This report was prepared as an account of work sponsored by an agency of the
United States Government. Neither the United States Government nor any agency thereof, nor
any of their employees, makes any warranty, express or implied, or assumes any legal liability
or responsibility for the accuracy, completeness, or usefulness of any information, apparatus,
product, or process disclosed, or represents that its use would not infringe privately owned
rights. Reference herein to any specific commercial product, process, or service by trade name,
trademark, manufacturer, or otherwise does not necessarily constitute or imply its
endorsement, recommendation, or favoring by the United States Government or any agency
thereof. The views and opinions of authors expressed herein do not necessarily state or reflect
those of the United States Government or any agency thereof.

\section*{Supporting Information}
Additional experiment details regarding neural network architecture, sampling, computation, and training, and additional validation experiments.

\bibliography{main.bib}

\providecommand{\latin}[1]{#1}
\makeatletter
\providecommand{\doi}
  {\begingroup\let\do\@makeother\dospecials
  \catcode`\{=1 \catcode`\}=2 \doi@aux}
\providecommand{\doi@aux}[1]{\endgroup\texttt{#1}}
\makeatother
\providecommand*\mcitethebibliography{\thebibliography}
\csname @ifundefined\endcsname{endmcitethebibliography}  {\let\endmcitethebibliography\endthebibliography}{}
\begin{mcitethebibliography}{47}
\providecommand*\natexlab[1]{#1}
\providecommand*\mciteSetBstSublistMode[1]{}
\providecommand*\mciteSetBstMaxWidthForm[2]{}
\providecommand*\mciteBstWouldAddEndPuncttrue
  {\def\EndOfBibitem{\unskip.}}
\providecommand*\mciteBstWouldAddEndPunctfalse
  {\let\EndOfBibitem\relax}
\providecommand*\mciteSetBstMidEndSepPunct[3]{}
\providecommand*\mciteSetBstSublistLabelBeginEnd[3]{}
\providecommand*\EndOfBibitem{}
\mciteSetBstSublistMode{f}
\mciteSetBstMaxWidthForm{subitem}{(\alph{mcitesubitemcount})}
\mciteSetBstSublistLabelBeginEnd
  {\mcitemaxwidthsubitemform\space}
  {\relax}
  {\relax}

\bibitem[Motta \latin{et~al.}(2020)Motta, Genovese, Ma, Cui, Sawaya, Chan, Chepiga, Helms, Jim\'enez-Hoyos, Millis, Ray, Ronca, Shi, Sorella, Stoudenmire, White, and Zhang]{motta2020ground}
Motta,~M.; Genovese,~C.; Ma,~F.; Cui,~Z.-H.; Sawaya,~R.; Chan,~G. K.-L.; Chepiga,~N.; Helms,~P.; Jim\'enez-Hoyos,~C.; Millis,~A.~J.; Ray,~U.; Ronca,~E.; Shi,~H.; Sorella,~S.; Stoudenmire,~E.~M. \latin{et~al.}  Ground-State Properties of the Hydrogen Chain: dimerization, Insulator-to-Metal Transition, and Magnetic Phases. \emph{Phys. Rev. X} 2020, \emph{10}, 031058, DOI: \doi{10.1103/PhysRevX.10.031058}\relax
\mciteBstWouldAddEndPuncttrue
\mciteSetBstMidEndSepPunct{\mcitedefaultmidpunct}
{\mcitedefaultendpunct}{\mcitedefaultseppunct}\relax
\EndOfBibitem
\bibitem[Williams \latin{et~al.}(2020)Williams, Yao, Li, Chen, Shi, Motta, Niu, Ray, Guo, Anderson, Li, Tran, Yeh, Mussard, Sharma, Bruneval, van Schilfgaarde, Booth, Chan, Zhang, Gull, Zgid, Millis, Umrigar, Wagner, and on~the Many-Electron~Problem]{williams-2020-direc-compar}
Williams,~K.~T.; Yao,~Y.; Li,~J.; Chen,~L.; Shi,~H.; Motta,~M.; Niu,~C.; Ray,~U.; Guo,~S.; Anderson,~R.~J.; Li,~J.; Tran,~L.~N.; Yeh,~C.-N.; Mussard,~B.; Sharma,~S. \latin{et~al.}  Direct Comparison of Many-Body Methods for Realistic Electronic {H}amiltonians. \emph{Physical Review X} 2020, \emph{10}, 011041, DOI: \doi{10.1103/physrevx.10.011041}\relax
\mciteBstWouldAddEndPuncttrue
\mciteSetBstMidEndSepPunct{\mcitedefaultmidpunct}
{\mcitedefaultendpunct}{\mcitedefaultseppunct}\relax
\EndOfBibitem
\bibitem[Martin(2020)]{martin2020electronic}
Martin,~R.~M. \emph{Electronic structure: basic theory and practical methods}; Cambridge university press, 2020\relax
\mciteBstWouldAddEndPuncttrue
\mciteSetBstMidEndSepPunct{\mcitedefaultmidpunct}
{\mcitedefaultendpunct}{\mcitedefaultseppunct}\relax
\EndOfBibitem
\bibitem[Smith and Burke(2017)Smith, and Burke]{smith2017interacting}
Smith,~J.~C.; Burke,~K. Interacting Electrons: theory and Computational Approaches. \emph{American Journal of Physics} 2017, \emph{85}, 636--637\relax
\mciteBstWouldAddEndPuncttrue
\mciteSetBstMidEndSepPunct{\mcitedefaultmidpunct}
{\mcitedefaultendpunct}{\mcitedefaultseppunct}\relax
\EndOfBibitem
\bibitem[Pfau \latin{et~al.}(2020)Pfau, Spencer, Matthews, and Foulkes]{pfau-2020-initio-solut}
Pfau,~D.; Spencer,~J.~S.; Matthews,~A. G. D.~G.; Foulkes,~W. M.~C. Ab Initio Solution of the Many-Electron {S}chr{\"o}dinger Equation With Deep Neural Networks. \emph{Physical Review Research} 2020, \emph{2}, 033429, DOI: \doi{10.1103/physrevresearch.2.033429}\relax
\mciteBstWouldAddEndPuncttrue
\mciteSetBstMidEndSepPunct{\mcitedefaultmidpunct}
{\mcitedefaultendpunct}{\mcitedefaultseppunct}\relax
\EndOfBibitem
\bibitem[Hermann \latin{et~al.}(2023)Hermann, Spencer, Choo, Mezzacapo, Foulkes, Pfau, Carleo, and No{\'e}]{hermann-2023-ab-initio}
Hermann,~J.; Spencer,~J.; Choo,~K.; Mezzacapo,~A.; Foulkes,~W. M.~C.; Pfau,~D.; Carleo,~G.; No{\'e},~F. Ab Initio Quantum Chemistry With Neural-Network Wavefunctions. \emph{Nature Reviews Chemistry} 2023, DOI: \doi{10.1038/s41570-023-00516-8}\relax
\mciteBstWouldAddEndPuncttrue
\mciteSetBstMidEndSepPunct{\mcitedefaultmidpunct}
{\mcitedefaultendpunct}{\mcitedefaultseppunct}\relax
\EndOfBibitem
\bibitem[Hermann \latin{et~al.}(2020)Hermann, Sch{\"a}tzle, and No{\'e}]{hermann-2020-deep-neural}
Hermann,~J.; Sch{\"a}tzle,~Z.; No{\'e},~F. Deep-Neural-Network Solution of the Electronic {S}chr{\"o}dinger Equation. \emph{Nature Chemistry} 2020, \emph{12}, 891--897, DOI: \doi{10.1038/s41557-020-0544-y}\relax
\mciteBstWouldAddEndPuncttrue
\mciteSetBstMidEndSepPunct{\mcitedefaultmidpunct}
{\mcitedefaultendpunct}{\mcitedefaultseppunct}\relax
\EndOfBibitem
\bibitem[Carleo and Troyer(2017)Carleo, and Troyer]{carleo-2017-solvin-quant}
Carleo,~G.; Troyer,~M. Solving the Quantum Many-Body Problem With Artificial Neural Networks. \emph{Science} 2017, \emph{355}, 602--606, DOI: \doi{10.1126/science.aag2302}\relax
\mciteBstWouldAddEndPuncttrue
\mciteSetBstMidEndSepPunct{\mcitedefaultmidpunct}
{\mcitedefaultendpunct}{\mcitedefaultseppunct}\relax
\EndOfBibitem
\bibitem[Luo and Clark(2019)Luo, and Clark]{luo-2019-backf-trans}
Luo,~D.; Clark,~B.~K. Backflow Transformations Via Neural Networks for Quantum Many-Body Wave Functions. \emph{Physical Review Letters} 2019, \emph{122}, 226401, DOI: \doi{10.1103/physrevlett.122.226401}\relax
\mciteBstWouldAddEndPuncttrue
\mciteSetBstMidEndSepPunct{\mcitedefaultmidpunct}
{\mcitedefaultendpunct}{\mcitedefaultseppunct}\relax
\EndOfBibitem
\bibitem[Li \latin{et~al.}(2022)Li, Li, and Chen]{li-2022-ab-initio}
Li,~X.; Li,~Z.; Chen,~J. Ab Initio Calculation of Real Solids Via Neural Network Ansatz. \emph{Nature Communications} 2022, \emph{13}, 7895, DOI: \doi{10.1038/s41467-022-35627-1}\relax
\mciteBstWouldAddEndPuncttrue
\mciteSetBstMidEndSepPunct{\mcitedefaultmidpunct}
{\mcitedefaultendpunct}{\mcitedefaultseppunct}\relax
\EndOfBibitem
\bibitem[Wilson \latin{et~al.}(2023)Wilson, Moroni, Holzmann, Gao, Wudarski, Vegge, and Bhowmik]{wilson-2023-neural-networ}
Wilson,~M.; Moroni,~S.; Holzmann,~M.; Gao,~N.; Wudarski,~F.; Vegge,~T.; Bhowmik,~A. Neural Network Ansatz for Periodic Wave Functions and the Homogeneous Electron Gas. \emph{Physical Review B} 2023, \emph{107}, 235139, DOI: \doi{10.1103/physrevb.107.235139}\relax
\mciteBstWouldAddEndPuncttrue
\mciteSetBstMidEndSepPunct{\mcitedefaultmidpunct}
{\mcitedefaultendpunct}{\mcitedefaultseppunct}\relax
\EndOfBibitem
\bibitem[Cassella \latin{et~al.}(2023)Cassella, Sutterud, Azadi, Drummond, Pfau, Spencer, and Foulkes]{cassella-2023-discov-quant}
Cassella,~G.; Sutterud,~H.; Azadi,~S.; Drummond,~N.~D.; Pfau,~D.; Spencer,~J.~S.; Foulkes,~W. M.~C. Discovering Quantum Phase Transitions With Fermionic Neural Networks. \emph{Physical Review Letters} 2023, \emph{130}, 036401, DOI: \doi{10.1103/physrevlett.130.036401}\relax
\mciteBstWouldAddEndPuncttrue
\mciteSetBstMidEndSepPunct{\mcitedefaultmidpunct}
{\mcitedefaultendpunct}{\mcitedefaultseppunct}\relax
\EndOfBibitem
\bibitem[Smith \latin{et~al.}(2024)Smith, Chen, Levy, Yang, Morales, and Zhang]{smith-2024-unified-variat}
Smith,~C.; Chen,~Y.; Levy,~R.; Yang,~Y.; Morales,~M.~A.; Zhang,~S. Unified Variational Approach Description of Ground-State Phases of the Two-Dimensional Electron Gas. \emph{Physical Review Letters} 2024, \emph{133}, 266504, DOI: \doi{10.1103/physrevlett.133.266504}\relax
\mciteBstWouldAddEndPuncttrue
\mciteSetBstMidEndSepPunct{\mcitedefaultmidpunct}
{\mcitedefaultendpunct}{\mcitedefaultseppunct}\relax
\EndOfBibitem
\bibitem[Yoshioka \latin{et~al.}(2021)Yoshioka, Mizukami, and Nori]{yoshioka-2021-solvin-quasip}
Yoshioka,~N.; Mizukami,~W.; Nori,~F. Solving Quasiparticle Band Spectra of Real Solids Using Neural-Network Quantum States. \emph{Communications Physics} 2021, \emph{4}, 106, DOI: \doi{10.1038/s42005-021-00609-0}\relax
\mciteBstWouldAddEndPuncttrue
\mciteSetBstMidEndSepPunct{\mcitedefaultmidpunct}
{\mcitedefaultendpunct}{\mcitedefaultseppunct}\relax
\EndOfBibitem
\bibitem[Scherbela \latin{et~al.}(2024)Scherbela, Gerard, and Grohs]{scherbela-2024-towar-trans}
Scherbela,~M.; Gerard,~L.; Grohs,~P. Towards a Transferable Fermionic Neural Wavefunction for Molecules. \emph{Nature Communications} 2024, \emph{15}, 120, DOI: \doi{10.1038/s41467-023-44216-9}\relax
\mciteBstWouldAddEndPuncttrue
\mciteSetBstMidEndSepPunct{\mcitedefaultmidpunct}
{\mcitedefaultendpunct}{\mcitedefaultseppunct}\relax
\EndOfBibitem
\bibitem[Linteau \latin{et~al.}(2025)Linteau, Moroni, Carleo, and Holzmann]{linteau-2025-univer-neural}
Linteau,~D.; Moroni,~S.; Carleo,~G.; Holzmann,~M. Universal Neural Wave Functions for High-Pressure Hydrogen. \emph{arXiv (Condensed Matter.Strongly Correlated Electrons)} 2025, DOI: \doi{10.48550/arXiv.2504.07062}\relax
\mciteBstWouldAddEndPuncttrue
\mciteSetBstMidEndSepPunct{\mcitedefaultmidpunct}
{\mcitedefaultendpunct}{\mcitedefaultseppunct}\relax
\EndOfBibitem
\bibitem[Xie \latin{et~al.}(2023)Xie, Li, Wang, Zhang, and Wang]{xie-2023-deep-variat}
Xie,~H.; Li,~Z.-H.; Wang,~H.; Zhang,~L.; Wang,~L. Deep Variational Free Energy Approach To Dense Hydrogen. \emph{Physical Review Letters} 2023, \emph{131}, 126501, DOI: \doi{10.1103/physrevlett.131.126501}\relax
\mciteBstWouldAddEndPuncttrue
\mciteSetBstMidEndSepPunct{\mcitedefaultmidpunct}
{\mcitedefaultendpunct}{\mcitedefaultseppunct}\relax
\EndOfBibitem
\bibitem[Pescia \latin{et~al.}(2024)Pescia, Nys, Kim, Lovato, and Carleo]{pescia-2024-messag-passin}
Pescia,~G.; Nys,~J.; Kim,~J.; Lovato,~A.; Carleo,~G. Message-Passing Neural Quantum States for the Homogeneous Electron Gas. \emph{Physical Review B} 2024, \emph{110}, 035108, DOI: \doi{10.1103/physrevb.110.035108}\relax
\mciteBstWouldAddEndPuncttrue
\mciteSetBstMidEndSepPunct{\mcitedefaultmidpunct}
{\mcitedefaultendpunct}{\mcitedefaultseppunct}\relax
\EndOfBibitem
\bibitem[Scherbela \latin{et~al.}(2025)Scherbela, Gao, Grohs, and G{\"u}nnemann]{scherbela-2025-accur-ab}
Scherbela,~M.; Gao,~N.; Grohs,~P.; G{\"u}nnemann,~S. Accurate Ab-Initio Neural-Network Solutions To Large-Scale Electronic Structure Problems. \emph{arXiv (Physics.Computational Physics)} 2025, DOI: \doi{10.48550/arXiv.2504.06087}\relax
\mciteBstWouldAddEndPuncttrue
\mciteSetBstMidEndSepPunct{\mcitedefaultmidpunct}
{\mcitedefaultendpunct}{\mcitedefaultseppunct}\relax
\EndOfBibitem
\bibitem[Gerard \latin{et~al.}(2022)Gerard, Scherbela, Marquetand, and Grohs]{NEURIPS2022_43089499}
Gerard,~L.; Scherbela,~M.; Marquetand,~P.; Grohs,~P. Gold-standard solutions to the {S}chr\"{o}dinger equation using deep learning: How much physics do we need? Advances in Neural Information Processing Systems. 2022; pp 10282--10294\relax
\mciteBstWouldAddEndPuncttrue
\mciteSetBstMidEndSepPunct{\mcitedefaultmidpunct}
{\mcitedefaultendpunct}{\mcitedefaultseppunct}\relax
\EndOfBibitem
\bibitem[Melton \latin{et~al.}(2016)Melton, Bennett, and Mitas]{melton2016quantum}
Melton,~C.~A.; Bennett,~M.~C.; Mitas,~L. {Quantum Monte Carlo with variable spins}. \emph{The Journal of Chemical Physics} 2016, \emph{144}, 244113, DOI: \doi{10.1063/1.4954726}\relax
\mciteBstWouldAddEndPuncttrue
\mciteSetBstMidEndSepPunct{\mcitedefaultmidpunct}
{\mcitedefaultendpunct}{\mcitedefaultseppunct}\relax
\EndOfBibitem
\bibitem[Melton \latin{et~al.}(2016)Melton, Zhu, Guo, Ambrosetti, Pederiva, and Mitas]{melton2016spin}
Melton,~C.~A.; Zhu,~M.; Guo,~S.; Ambrosetti,~A.; Pederiva,~F.; Mitas,~L. Spin-orbit interactions in electronic structure quantum {M}onte {C}arlo methods. \emph{Phys. Rev. A} 2016, \emph{93}, 042502, DOI: \doi{10.1103/PhysRevA.93.042502}\relax
\mciteBstWouldAddEndPuncttrue
\mciteSetBstMidEndSepPunct{\mcitedefaultmidpunct}
{\mcitedefaultendpunct}{\mcitedefaultseppunct}\relax
\EndOfBibitem
\bibitem[Schmidt and Fantoni(1999)Schmidt, and Fantoni]{schmidt-1999-quant-monte}
Schmidt,~K.; Fantoni,~S. A Quantum {M}onte {C}arlo Method for Nucleon Systems. \emph{Physics Letters B} 1999, \emph{446}, 99--103, DOI: \doi{10.1016/s0370-2693(98)01522-6}\relax
\mciteBstWouldAddEndPuncttrue
\mciteSetBstMidEndSepPunct{\mcitedefaultmidpunct}
{\mcitedefaultendpunct}{\mcitedefaultseppunct}\relax
\EndOfBibitem
\bibitem[Lonardoni \latin{et~al.}(2018)Lonardoni, Gandolfi, Lynn, Petrie, Carlson, Schmidt, and Schwenk]{lonardoni-2018-auxil-field}
Lonardoni,~D.; Gandolfi,~S.; Lynn,~J.~E.; Petrie,~C.; Carlson,~J.; Schmidt,~K.~E.; Schwenk,~A. Auxiliary Field Diffusion {M}onte {C}arlo Calculations of Light and Medium-Mass Nuclei With Local Chiral Interactions. \emph{Physical Review C} 2018, \emph{97}, 044318, DOI: \doi{10.1103/physrevc.97.044318}\relax
\mciteBstWouldAddEndPuncttrue
\mciteSetBstMidEndSepPunct{\mcitedefaultmidpunct}
{\mcitedefaultendpunct}{\mcitedefaultseppunct}\relax
\EndOfBibitem
\bibitem[Huang \latin{et~al.}(1998)Huang, Filippi, and Umrigar]{huang-1998-spin-contam}
Huang,~C.-J.; Filippi,~C.; Umrigar,~C.~J. Spin Contamination in Quantum Monte Carlo Wave Functions. \emph{The Journal of Chemical Physics} 1998, \emph{108}, 8838--8847, DOI: \doi{10.1063/1.476330}\relax
\mciteBstWouldAddEndPuncttrue
\mciteSetBstMidEndSepPunct{\mcitedefaultmidpunct}
{\mcitedefaultendpunct}{\mcitedefaultseppunct}\relax
\EndOfBibitem
\bibitem[Kim \latin{et~al.}(2024)Kim, Pescia, Fore, Nys, Carleo, Gandolfi, Hjorth-Jensen, and Lovato]{kim-2024-neural-networ}
Kim,~J.; Pescia,~G.; Fore,~B.; Nys,~J.; Carleo,~G.; Gandolfi,~S.; Hjorth-Jensen,~M.; Lovato,~A. Neural-Network Quantum States for Ultra-Cold {F}ermi Gases. \emph{Communications Physics} 2024, \emph{7}, 148, DOI: \doi{10.1038/s42005-024-01613-w}\relax
\mciteBstWouldAddEndPuncttrue
\mciteSetBstMidEndSepPunct{\mcitedefaultmidpunct}
{\mcitedefaultendpunct}{\mcitedefaultseppunct}\relax
\EndOfBibitem
\bibitem[Adams \latin{et~al.}(2021)Adams, Carleo, Lovato, and Rocco]{adams-2021-variat-monte}
Adams,~C.; Carleo,~G.; Lovato,~A.; Rocco,~N. Variational {M}onte {C}arlo Calculations of ${A} \leq 4$ Nuclei With an Artificial Neural-Network Correlator Ansatz. \emph{Physical Review Letters} 2021, \emph{127}, 022502, DOI: \doi{10.1103/physrevlett.127.022502}\relax
\mciteBstWouldAddEndPuncttrue
\mciteSetBstMidEndSepPunct{\mcitedefaultmidpunct}
{\mcitedefaultendpunct}{\mcitedefaultseppunct}\relax
\EndOfBibitem
\bibitem[Luo \latin{et~al.}(2025)Luo, Zaklama, and Fu]{luo-2025-solvin-fract}
Luo,~D.; Zaklama,~T.; Fu,~L. Solving Fractional Electron States in Twisted {M}o{T}e$_2$ With Deep Neural Network. \emph{arXiv (Condensed Matter.Strongly Correlated Electrons)} 2025, DOI: \doi{10.48550/arXiv.2503.13585}\relax
\mciteBstWouldAddEndPuncttrue
\mciteSetBstMidEndSepPunct{\mcitedefaultmidpunct}
{\mcitedefaultendpunct}{\mcitedefaultseppunct}\relax
\EndOfBibitem
\bibitem[Li \latin{et~al.}(2024)Li, Lu, Li, Wen, Li, Wang, Chen, and Ren]{li-2024-spin-symmet}
Li,~Z.; Lu,~Z.; Li,~R.; Wen,~X.; Li,~X.; Wang,~L.; Chen,~J.; Ren,~W. Spin-Symmetry-Enforced Solution of the Many-Body {S}chr{\"o}dinger Equation With a Deep Neural Network. \emph{Nature Computational Science} 2024, \emph{4}, 910--919, DOI: \doi{10.1038/s43588-024-00730-4}\relax
\mciteBstWouldAddEndPuncttrue
\mciteSetBstMidEndSepPunct{\mcitedefaultmidpunct}
{\mcitedefaultendpunct}{\mcitedefaultseppunct}\relax
\EndOfBibitem
\bibitem[Szab{\'o} \latin{et~al.}(2024)Szab{\'o}, Sch{\"a}tzle, Entwistle, and No{\'e}]{szabo-2024-improv-penal}
Szab{\'o},~P.~B.; Sch{\"a}tzle,~Z.; Entwistle,~M.~T.; No{\'e},~F. An Improved Penalty-Based Excited-State Variational {M}onte {C}arlo Approach With Deep-Learning Ansatzes. \emph{Journal of Chemical Theory and Computation} 2024, DOI: \doi{10.1021/acs.jctc.4c00678}\relax
\mciteBstWouldAddEndPuncttrue
\mciteSetBstMidEndSepPunct{\mcitedefaultmidpunct}
{\mcitedefaultendpunct}{\mcitedefaultseppunct}\relax
\EndOfBibitem
\bibitem[Lin \latin{et~al.}(2023)Lin, Goldshlager, and Lin]{lin-2023-explic-antis}
Lin,~J.; Goldshlager,~G.; Lin,~L. Explicitly Antisymmetrized Neural Network Layers for Variational {M}onte {C}arlo Simulation. \emph{Journal of Computational Physics} 2023, \emph{474}, 111765, DOI: \doi{10.1016/j.jcp.2022.111765}\relax
\mciteBstWouldAddEndPuncttrue
\mciteSetBstMidEndSepPunct{\mcitedefaultmidpunct}
{\mcitedefaultendpunct}{\mcitedefaultseppunct}\relax
\EndOfBibitem
\bibitem[Sch{\"a}tzle \latin{et~al.}(2023)Sch{\"a}tzle, Szab{\'o}, Mezera, Hermann, and No{\'e}]{schaetzle-2023-deepq}
Sch{\"a}tzle,~Z.; Szab{\'o},~P.~B.; Mezera,~M.; Hermann,~J.; No{\'e},~F. Deepqmc: an Open-Source Software Suite for Variational Optimization of Deep-Learning Molecular Wave Functions. \emph{The Journal of Chemical Physics} 2023, \emph{159}, DOI: \doi{10.1063/5.0157512}\relax
\mciteBstWouldAddEndPuncttrue
\mciteSetBstMidEndSepPunct{\mcitedefaultmidpunct}
{\mcitedefaultendpunct}{\mcitedefaultseppunct}\relax
\EndOfBibitem
\bibitem[Pauli(1927)]{pauli-1927-zur-quant}
Pauli,~W. Zur Quantenmechanik Des Magnetischen Elektrons. \emph{Zeitschrift f{\"u}r Physik} 1927, \emph{43}, 601--623, DOI: \doi{10.1007/bf01397326}\relax
\mciteBstWouldAddEndPuncttrue
\mciteSetBstMidEndSepPunct{\mcitedefaultmidpunct}
{\mcitedefaultendpunct}{\mcitedefaultseppunct}\relax
\EndOfBibitem
\bibitem[Glehn \latin{et~al.}(2022)Glehn, Spencer, and Pfau]{glehn-2022-self-atten}
Glehn,~I.~v.; Spencer,~J.~S.; Pfau,~D. A Self-Attention Ansatz for Ab-Initio Quantum Chemistry. \emph{arXiv (Physics.Chemical Physics)} 2022, DOI: \doi{10.48550/arXiv.2211.13672}\relax
\mciteBstWouldAddEndPuncttrue
\mciteSetBstMidEndSepPunct{\mcitedefaultmidpunct}
{\mcitedefaultendpunct}{\mcitedefaultseppunct}\relax
\EndOfBibitem
\bibitem[Foulkes \latin{et~al.}(2001)Foulkes, Mitas, Needs, and Rajagopal]{foulkes-2001-quant-monte}
Foulkes,~W. M.~C.; Mitas,~L.; Needs,~R.~J.; Rajagopal,~G. Quantum {M}onte {C}arlo Simulations of Solids. \emph{Reviews of Modern Physics} 2001, \emph{73}, 33--83, DOI: \doi{10.1103/revmodphys.73.33}\relax
\mciteBstWouldAddEndPuncttrue
\mciteSetBstMidEndSepPunct{\mcitedefaultmidpunct}
{\mcitedefaultendpunct}{\mcitedefaultseppunct}\relax
\EndOfBibitem
\bibitem[Fisher(1953)]{fisher-1953-dispersion}
Fisher,~R.~A. Dispersion on a sphere. \emph{Proceedings of the Royal Society of London. Series A. Mathematical and Physical Sciences} 1953, \emph{217}, 295--305, DOI: \doi{10.1098/rspa.1953.0064}\relax
\mciteBstWouldAddEndPuncttrue
\mciteSetBstMidEndSepPunct{\mcitedefaultmidpunct}
{\mcitedefaultendpunct}{\mcitedefaultseppunct}\relax
\EndOfBibitem
\bibitem[Zhan \latin{et~al.}(2025)Zhan, Wheeler, Goldshlager, Ertekin, Adams, and Wagner]{zhan_2025_data}
Zhan,~N.; Wheeler,~W.~A.; Goldshlager,~G.; Ertekin,~E.; Adams,~R.~P.; Wagner,~L.~K. Data: Expressivity of determinantal anzatzes for neural network wave functions. 2025; \url{https://doi.org/10.5281/zenodo.16506718}\relax
\mciteBstWouldAddEndPuncttrue
\mciteSetBstMidEndSepPunct{\mcitedefaultmidpunct}
{\mcitedefaultendpunct}{\mcitedefaultseppunct}\relax
\EndOfBibitem
\bibitem[Frankle and Carbin(2018)Frankle, and Carbin]{frankle-2018-lotter-ticket-hypot}
Frankle,~J.; Carbin,~M. The Lottery Ticket Hypothesis: Finding Sparse, Trainable Neural Networks. \emph{arXiv (Computer Science.Machine Learning)} 2018, DOI: \doi{10.48550/arXiv.1803.03635}\relax
\mciteBstWouldAddEndPuncttrue
\mciteSetBstMidEndSepPunct{\mcitedefaultmidpunct}
{\mcitedefaultendpunct}{\mcitedefaultseppunct}\relax
\EndOfBibitem
\bibitem[Lawrence \latin{et~al.}(1997)Lawrence, Giles, and Tsoi]{lawrence1997lessons}
Lawrence,~S.; Giles,~C.~L.; Tsoi,~A.~C. Lessons in neural network training: Overfitting may be harder than expected. Aaai/iaai. 1997; pp 540--545\relax
\mciteBstWouldAddEndPuncttrue
\mciteSetBstMidEndSepPunct{\mcitedefaultmidpunct}
{\mcitedefaultendpunct}{\mcitedefaultseppunct}\relax
\EndOfBibitem
\bibitem[Zou and Gu(2019)Zou, and Gu]{NEURIPS2019_6a61d423}
Zou,~D.; Gu,~Q. An Improved Analysis of Training Over-parameterized Deep Neural Networks. Advances in Neural Information Processing Systems. 2019\relax
\mciteBstWouldAddEndPuncttrue
\mciteSetBstMidEndSepPunct{\mcitedefaultmidpunct}
{\mcitedefaultendpunct}{\mcitedefaultseppunct}\relax
\EndOfBibitem
\bibitem[Allen-Zhu \latin{et~al.}(2019)Allen-Zhu, Li, and Song]{pmlr-v97-allen-zhu19a}
Allen-Zhu,~Z.; Li,~Y.; Song,~Z. A Convergence Theory for Deep Learning via Over-Parameterization. Proceedings of the 36th International Conference on Machine Learning. 2019; pp 242--252\relax
\mciteBstWouldAddEndPuncttrue
\mciteSetBstMidEndSepPunct{\mcitedefaultmidpunct}
{\mcitedefaultendpunct}{\mcitedefaultseppunct}\relax
\EndOfBibitem
\bibitem[Du \latin{et~al.}(2019)Du, Lee, Li, Wang, and Zhai]{pmlr-v97-du19c}
Du,~S.; Lee,~J.; Li,~H.; Wang,~L.; Zhai,~X. Gradient Descent Finds Global Minima of Deep Neural Networks. Proceedings of the 36th International Conference on Machine Learning. 2019; pp 1675--1685\relax
\mciteBstWouldAddEndPuncttrue
\mciteSetBstMidEndSepPunct{\mcitedefaultmidpunct}
{\mcitedefaultendpunct}{\mcitedefaultseppunct}\relax
\EndOfBibitem
\bibitem[Jim{\'e}nez-Hoyos \latin{et~al.}(2014)Jim{\'e}nez-Hoyos, Rodr{\'i}guez-Guzm{\'a}n, and Scuseria]{jimenez-hoyos-2014-polyr-charac}
Jim{\'e}nez-Hoyos,~C.~A.; Rodr{\'i}guez-Guzm{\'a}n,~R.; Scuseria,~G.~E. Polyradical Character and Spin Frustration in Fullerene Molecules: an Ab Initio Non-Collinear Hartree-Fock Study. \emph{The Journal of Physical Chemistry A} 2014, \emph{118}, 9925--9940, DOI: \doi{10.1021/jp508383z}\relax
\mciteBstWouldAddEndPuncttrue
\mciteSetBstMidEndSepPunct{\mcitedefaultmidpunct}
{\mcitedefaultendpunct}{\mcitedefaultseppunct}\relax
\EndOfBibitem
\bibitem[Ambrosetti \latin{et~al.}(2009)Ambrosetti, Pederiva, Lipparini, and Gandolfi]{ambrosetti-2009-quant-monte}
Ambrosetti,~A.; Pederiva,~F.; Lipparini,~E.; Gandolfi,~S. Quantum {M}onte {C}arlo Study of the Two-Dimensional Electron Gas in Presence of {R}ashba Interaction. \emph{Physical Review B} 2009, \emph{80}, 125306, DOI: \doi{10.1103/physrevb.80.125306}\relax
\mciteBstWouldAddEndPuncttrue
\mciteSetBstMidEndSepPunct{\mcitedefaultmidpunct}
{\mcitedefaultendpunct}{\mcitedefaultseppunct}\relax
\EndOfBibitem
\bibitem[Goldshlager \latin{et~al.}(2024)Goldshlager, Abrahamsen, and Lin]{goldshlager-2024-kaczm-inspir}
Goldshlager,~G.; Abrahamsen,~N.; Lin,~L. A {K}aczmarz-Inspired Approach To Accelerate the Optimization of Neural Network Wavefunctions. \emph{Journal of Computational Physics} 2024, \emph{516}, 113351, DOI: \doi{10.1016/j.jcp.2024.113351}\relax
\mciteBstWouldAddEndPuncttrue
\mciteSetBstMidEndSepPunct{\mcitedefaultmidpunct}
{\mcitedefaultendpunct}{\mcitedefaultseppunct}\relax
\EndOfBibitem
\bibitem[Pakyuz-Charrier \latin{et~al.}(2018)Pakyuz-Charrier, Lindsay, Ogarko, Giraud, and Jessell]{pakyuz-charrier-2018-monte-carlo}
Pakyuz-Charrier,~E.; Lindsay,~M.; Ogarko,~V.; Giraud,~J.; Jessell,~M. Monte {C}arlo Simulation for Uncertainty Estimation on Structural Data in Implicit 3-d Geological Modeling, a Guide for Disturbance Distribution Selection and Parameterization. \emph{Solid Earth} 2018, \emph{9}, 385--402, DOI: \doi{10.5194/se-9-385-2018}\relax
\mciteBstWouldAddEndPuncttrue
\mciteSetBstMidEndSepPunct{\mcitedefaultmidpunct}
{\mcitedefaultendpunct}{\mcitedefaultseppunct}\relax
\EndOfBibitem
\bibitem[Wood(1994)]{wood-1994-simul-von}
Wood,~A.~T. Simulation of the {V}on {M}ises {F}isher Distribution. \emph{Communications in Statistics - Simulation and Computation} 1994, \emph{23}, 157--164, DOI: \doi{10.1080/03610919408813161}\relax
\mciteBstWouldAddEndPuncttrue
\mciteSetBstMidEndSepPunct{\mcitedefaultmidpunct}
{\mcitedefaultendpunct}{\mcitedefaultseppunct}\relax
\EndOfBibitem
\end{mcitethebibliography}

\clearpage
\onecolumngrid

\begingroup  

\renewcommand\thesection{\Roman{section}} 
\setcounter{section}{0}  
\setcounter{figure}{0}  
\makeatletter
\renewcommand{\thefigure}{S\@arabic\c@figure}
\renewcommand{\thetable}{S\@arabic\c@table}

\section*{Supporting Information: Expressivity of determinantal ansatzes for neural network wave functions}
\addcontentsline{toc}{section}{Supporting Information}  

\title{Supporting Information: Expressivity of determinantal anzatzes for neural network wave functions}

\maketitle

\section{Network architectural details}

In our work, we adapt FermiNet's architecture for variable spins \cite{pfau-2020-initio-solut}. Fig.~\ref{fig-network-diagram} shows the overall wave function and network architecture. FermiNet has two neural networks, one for electron-nuclear feature input (single stream) and one for electron-electron (double stream). The final output from the single stream MLPs are multiplied by envelopes and a possible phase to form $\phi_{j, \{\uparrow, \downarrow\}}$. Then we multiply with the spinors $s_i$ and form the determinant. 

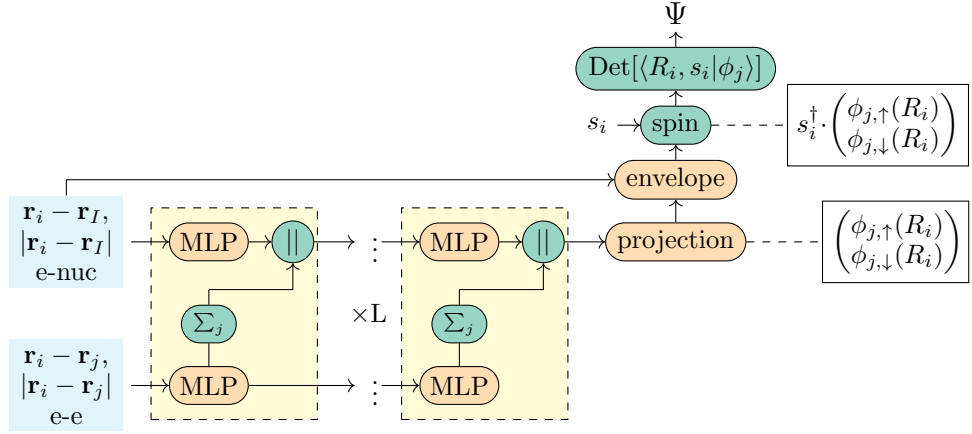
\begin{figure}[ht!]
\begin{tikzpicture}[learn/.style={rounded rectangle, draw, fill=orange!90!yellow!30, inner sep=1.0mm},
    fixed/.style={rounded rectangle, draw, fill=cyan!30!green!40, inner sep=1.0mm},
box/.style={rectangle, draw, dashed, minimum height=2.8cm, minimum width=2.2cm, fill=yellow!20},
 MLP/.style={learn, minimum width=1cm, minimum height=5mm},inp/.style={rectangle, fill=cyan!10, inner sep=1.0mm},spinornet/.style={rectangle, draw},
    arrows={-Computer Modern Rightarrow [width=10pt]}]

\node(box) [box]{};

\node(one) [MLP, below right=2mm and 5mm of box.north west] { MLP};
\node(two) [MLP, above right= 2mm and 5mm of box.south west] { MLP};
\node(sum) [fixed,  above=3mm of two.north] {\scriptsize $\sum_j$};
\node(inconcat) [coordinate, above=2mm of sum.north]{};

\node(concat) [fixed, right=3mm  of one.east]{ $||$};

\draw[->] (one.east) -- ++(3mm, 0mm);
\draw[-] (sum) -- (inconcat);
\draw[->] (inconcat) -| (concat.south);
\draw[-] (two) -- (sum);

\node(leftone) [coordinate, left=5mm of one]{};
\node(lefttwo) [coordinate, left=5mm of two]{};

\draw[->] (leftone) -- (one);
\draw[->] (lefttwo) -- (two);

\node (tiptwo) [coordinate, right=14mm of two]{};
\node (outone) at (concat.east -| one.east)[coordinate]{};

\node (tipone) at (outone -| tiptwo)[coordinate]{};

\draw[->] (concat.east) -- (tipone);

\node(iI)[inp, left=1mm of leftone]{\parbox{13mm}{\centering $\mathbf{r}_i - \mathbf{r}_I$,\\ $|\mathbf{r}_i - \mathbf{r}_I|$\\e-nuc}};
\node(ij)[inp, left=1mm of lefttwo]{\parbox{13mm}{\centering $\mathbf{r}_i - \mathbf{r}_j$,\\ $|\mathbf{r}_i - \mathbf{r}_j|$\\ e-e}};

\node (dotsone) [right  =1mm of tipone]{$\vdots$};
\node (dotstwo) [right=1mm  of tiptwo]{$\vdots$};
\node (timesL) [right=3mm of box.east]{$\times$L};
\draw[->] (two) -- (tiptwo);

\node(box) [box, right=11mm of box]{};

\node(one) [MLP, below right=2mm and 5mm of box.north west] { MLP};
\node(two) [MLP, above right= 2mm and 5mm of box.south west] { MLP};
\node(sum) [fixed,  above=3mm of two.north] {\scriptsize $\sum_j$};
\node(inconcat) [coordinate, above=2mm of sum.north]{};

\node(concat) [fixed, right=3mm  of one.east]{ $||$};

\draw[->] (one.east) -- ++(3mm, 0mm);
\draw[-] (sum) -- (inconcat);
\draw[->] (inconcat) -| (concat.south);
\draw[-] (two) -- (sum);

\node(leftone) [coordinate, left=5mm of one]{};
\node(lefttwo) [coordinate, left=5mm of two]{};

\draw[->] (leftone) -- (one);
\draw[->] (lefttwo) -- (two);

\node (tiptwo) [coordinate, right=14mm of two]{};
\node (outone) at (concat.east -| one.east)[coordinate]{};

\node (tipone) at (outone -| tiptwo)[coordinate]{};

\draw[->] (concat.east) -- (tipone);

\node(projection)[learn, right=0mm of tipone]{projection};
\node(envelope)[learn, above=3mm of projection]{envelope};

\draw[->] (iI.north)  ++(0mm, 0mm) |- (envelope);

\node(spin)[fixed, above=2mm of envelope]{spin};
\node(sigma)[left=3mm of spin]{$s_i$};
\draw[->] (projection) -- (envelope);
\draw[->] (envelope) -- (spin);
\draw[->] (sigma) -- (spin);

\node(det)[fixed, above=2mm of spin]{Det$[\braket{R_i, s_i |\phi_j}]$};

\node(psi)[above=2mm of det, inner sep=1mm]{\large $\Psi$};
\draw[->] (spin) -- (det);
\draw[->] (det) -- (psi);


\node(spindescr) [spinornet, right =10mm  of spin]{\parbox{2.16cm}{
$s_i^\dagger \cdot \begin{pmatrix}
           \phi_{j,\uparrow}(R_i) \\
           \phi_{j,\downarrow}(R_i)
         \end{pmatrix}$}};


\node(projdescr) [spinornet, right =10mm  of projection]{\parbox{1.7cm}{
$\begin{pmatrix}
           \phi_{j,\uparrow}(R_i) \\
           \phi_{j,\downarrow}(R_i)
         \end{pmatrix}$}};


\begin{scope}[-]
  \path
(spin) edge[dashed] (spindescr);
\path
(projection) edge[dashed] (projdescr);
\end{scope}

\end{tikzpicture}
\caption{Our ansatz introduces the dot product of spin-orbitals $\phi_{\uparrow,\downarrow}$ with spinor $s$. MLP: multilayer perceptron. $\| $ denotes concatenation. Figure based on \cite{glehn-2022-self-atten}. 
}
\label{fig-network-diagram}
\end{figure}
         
         We denote the concatenation of the electron-nuclear features as $\mathbf{h}_i^0$ and the electron-electron features as $\mathbf{h}_{ij}^0$. Assuming the outputs of the single electron network at layer $l$ are $\mathbf{h}_i^l$ and outputs of the double electron network are $\mathbf{h}_{ij}^l$, the input to the $l+1$ layer of the single stream network, for electron $i$, is 
         \begin{equation}
\begin{split}
\left( \mathbf{h}_i^l, \frac{1}{n} \sum_{j=1} \mathbf{h}_j^l, \frac{1}{n}\sum_{j=1}  \mathbf{h}_{ij}^l  \right) = (\mathbf{h}_i^l, \mathbf{g}^l, \mathbf{g}_i^l) = \mathbf{f}_i^l
\end{split}
\end{equation} where $n$ is number of electrons. Since we include spin information in the double electron features, we average over all electrons instead of splitting averages by spin as the original FermiNet. The NNs' layer computations are

\begin{equation}
\begin{split}
\mathbf{h}_i^{l+1} = \tanh(\mathbf{W}^l\mathbf{f}_i^l + \mathbf{b}^l) \\
\mathbf{h}_{ij}^{l+1} = \tanh(\mathbf{V}^l\mathbf{h}_{ij}^l + \mathbf{c}^l).
\end{split}
\end{equation}

The final output is linearly transformed, multiplied by an envelope, and reshaped to the orbitals $\phi_{j,\{\uparrow, \downarrow\}}(R_i)$. 

\begin{equation}
\begin{split}
\phi_{j,\{\uparrow, \downarrow\}}(R_i) = (\mathbf{w}_{j,\{\uparrow, \downarrow\}} \cdot \mathbf{h}_{i}^L+ b_{j,\{\uparrow, \downarrow\}}) \cdot \text{env}_{j,\{\uparrow, \downarrow\}}(\mathbf{r}_i).
\end{split}
\end{equation}

\subsection{System-specific changes}
\label{appendix-system-specific-changes}

\begin{table*}[ht]
  \caption{System specific modifications. $s(\cdot)$ indicates periodic transform. $\| \cdot\|_p$ indicates modified periodic approximate size (Eq.~\ref{eq:mod-periodic-norm}).}
  \label{table-solid-mol-heg-summary}
  \centering
  \scalebox{0.9}{
  \begin{tabular}{lll}
    \toprule
 System  &NN Input Features  & Envelopes  \\
\midrule
 Molecules  & $\{\mathbf{r}_i -\mathbf{r}_I \}, \{|\mathbf{r}_i -\mathbf{r}_I |\}, \{\mathbf{r}_i - \mathbf{r}_j \}, \{|\mathbf{r}_i -\mathbf{r}_j | \}$ & Atomic (Eq.~\ref{eq:atom-env})\\
 Solids & $\{s(\mathbf{r}_i -\mathbf{r}_I)\}, \{\|\mathbf{r}_i -\mathbf{r}_I \|_p\}, \{s(\mathbf{r}_i -\mathbf{r}_j)\}, \{\|\mathbf{r}_i -\mathbf{r}_j \|_p\}$ & Atomic and Phase (Eqs.~\ref{eq:atom-env}, \ref{eq:phase-env})\\
 HEG & $\{s(\mathbf{r}_i)\}, \{s(\mathbf{r}_i -\mathbf{r}_j)\}, \{\|\mathbf{r}_i -\mathbf{r}_j \|_p\}$ & Phase (Eq.~\ref{eq:phase-env}), Rashba (Eq.~\ref{eq:rashba-env})\\
    \bottomrule
  \end{tabular}
  }
\end{table*}

Our ansatz is general for molecules, solids, and HEG. Molecules use open boundary conditions, while solids and the HEG have periodic boundaries. Further, in molecules and solids the attractive potential for the electrons is generated by atomic nuclei,  while in the HEG, it comes from a uniform positive background. Hence, some modifications to the input features and envelopes are required depending on the system. Table \ref{table-solid-mol-heg-summary} summarizes the changes, and Sec.~\ref{sec-envelopes} describes the envelopes. In addition, we allow orbitals to be complex, predicting their real and imaginary components.

For periodic systems, we simulate a finite size cell with lattice vectors $\{\mathbf{a}_1, \mathbf{a}_2, \mathbf{a}_3 \}$, called the simulation cell, that is infinitely tiled through space. The simulation cell itself may be a tiling of smaller ``primitive cells". The wave function must satisfy $\Psi(...,\mathbf{r}_i + \mathbf{a}, ...) = \Psi(...,\mathbf{r}_i,...)$ for each $\mathbf{a}$. To satisfy this constraint, we modify the input features to the NN as proposed by \cite{cassella-2023-discov-quant}.  

In the periodic transform, we write a vector $\mathbf{r} := s_1 \mathbf{a}_1 + s_2 \mathbf{a}_2 + s_3 \mathbf{a}_3$ and transform $s_i \rightarrow [\sin(2\pi s_i), \cos(2 \pi s_i)]$ and use the approximate size in place of the Euclidean norm
         \begin{equation}\label{eq:mod-periodic-norm}
\| \mathbf{r}\|^2_p =\sum_{ij} [1 - \cos(2 \pi s_i)]S_{ij}[1 - \cos(2 \pi s_j)] 
+ \sin(2 \pi s_i)S_{ij}\sin(2 \pi s_j),
\end{equation} where $S_{ij} = \mathbf{a}_i \cdot \mathbf{a}_j$ is a scaling factor to approximate real space. We transform electron-electron features with the simulation lattice and electron-nuclear features with the primitive lattice, since atoms of the same primitive cell coordinate are equivalent. The approximate size $\| \cdot \|_p$ is periodic with the lattice, appears like absolute value when $\mathbf{r}\rightarrow0$, and is smooth elsewhere. The periodicity and smoothness are necessary for the wave function, and the sharpness (or cusp) near $\mathbf{r}=0$ helps minimize the loss. For solids, all transformed features are included; for the HEG, we exclude $\{\|\mathbf{r}_i - \mathbf{r}_I\|_p\}$ and include the transformed $\{\mathbf{r}_i - \mathbf{r}_I\}$ with $\mathbf{r}_I$ being the origin.

\subsection{Envelopes}
\label{sec-envelopes}
Envelopes are used in neural network ansatzes to improve solution convergence. The atomic envelope makes the wave function probability small when electrons are far from nuclei, and is used for molecules and periodic atomic systems (solids). The atomic envelope is a Gaussian around the nuclei positions 

\begin{equation} \label{eq:atom-env}
\begin{split}
\text{env}_{j,\{\uparrow, \downarrow\}}(\mathbf{r}_i) = \sum_I \pi^I_{j,\{\uparrow, \downarrow\}} \exp(-|\mathbf{\Sigma}^I_{j,\{\uparrow, \downarrow\}} (\mathbf{r}_i - \mathbf{r}_I) |)
\end{split}
\end{equation}

where $\pi^I_{j,\{\uparrow, \downarrow\}}$ and $\mathbf{\Sigma}^I_{j,\{\uparrow, \downarrow\}}$ are learnable parameters. The envelope for atomic solids includes Eq.~\ref{eq:atom-env} with $\pi^I_{j,\{\uparrow, \downarrow\}}$ and $\mathbf{\Sigma}^I_{j,\{\uparrow, \downarrow\}}$ shared across atoms of the primitive cell, the periodic transform of $\mathbf{r}_i - \mathbf{r}_I$ as input, 
and an additional phase envelope

\begin{equation} \label{eq:phase-env}
\begin{split}
\text{env}_{j,\{\uparrow, \downarrow\}}(\mathbf{r}_i) = \exp(i \mathbf{r}_i \mathbf{k}_j)
\end{split}
\end{equation}

where $\mathbf{k}_j$ are the primitive cell reciprocal lattice vectors. The HEG envelope includes Eq.~\ref{eq:phase-env} only. 

For 2DEG with Rashba, we use the Rashba envelope, based on the eigenstates of the noninteracting Rashba Hamiltonian.

\begin{equation} \label{eq:rashba-env}
\begin{split}
\text{env}_{j,\{\uparrow, \downarrow\}}(\mathbf{r}_i) =  \exp(i \mathbf{r}_i \mathbf{k}_j) \begin{pmatrix} \pm \frac{k_{j,y}+ik_{j,x}}{k_j} \\ 1 \end{pmatrix}
\end{split}
\end{equation}

The Rashba envelopes are chosen with $n_\uparrow$ spin-orbitals with the positive coefficient $\frac{k_y + i k_x}{k}$ and $n_\downarrow$ spin-orbitals with the negative coefficient. 

All systems can be multiplied with a collinear envelope that zeroes out corresponding spin-orbital components to make the ansatzes collinear. 

\section{Sampling details}

We implemented Metropolis-Hastings (MH) with Gaussian proposal and Metropolis-adjusted Langevin algorithm (MALA) for the positions moves. For MALA algorithm, the gradient magnitude is limited to $\leq 0.1$. We also implemented discrete and continuous spin sampling. For spin sampling, we move spins one electron at a time and propose a random ordering of the electrons at each MH step. During discrete sampling, electron spins are either $\begin{bmatrix} 1 \\ 0\end{bmatrix}$ or $\begin{bmatrix} 0 \\ 1\end{bmatrix}$, and the proposal is to flip the electron's current spin. During continuous sampling, the electron spin is a normalized complex two-vector (anywhere on Bloch sphere), and the proposal is a von-Mises Fisher (VMF) distribution on the sphere centered around the electron's current spin \cite{fisher-1953-dispersion}. 

The VMF distribution is a probability distribution on the ($p-1$)-sphere in $\mathbb{R}^p$. We use $p=3$ for $S^2$ sphere. The probability density function is 

\begin{equation}
f(x|\mu, \kappa) = C_p(\kappa) \exp(\kappa \mu^T x) \end{equation} where $\kappa \geq 0$, $||\mu||=1$, and

\begin{equation}
C_p(\kappa) = \frac{\kappa^{p/2-1}}{(2\pi)^{p/2}I_{p/2-1}(\kappa)}
\end{equation}

where $I_v$ is the modified Bessel function of the first kind at order $v$. The parameter $\mu$ is directional mean, and $\kappa$ is concentration. For $\kappa=0$, the distribution is uniform on the sphere, and as $\kappa$ increases, the distribution becomes peaked around the mean. 

The unit vectors on $S^2$ can be defined with polar angle $\theta$ and azimuthal angle $\phi$. We sample $\theta$ and $\phi$ from the VMF distribution with mean at $[0, 0, 1]$. Then we rotate the current spin based on the sampled $\theta$ and $\phi$. These are sampled as follows, from \cite{pakyuz-charrier-2018-monte-carlo,wood-1994-simul-von}.

\begin{equation}
\phi \sim U(0, 2\pi)
\end{equation}

\begin{equation}
\cos \theta = 1 + \frac{1}{\kappa}\left(\log \xi + \log \left(1 - \frac{\xi - 1}{\xi}\right)\exp^{-2\kappa}\right)
\end{equation}

\begin{equation}
\xi \sim U(0, 1)
\end{equation}

where $U(a, b)$ is the uniform distribution. For our experiments, we set $\kappa=1.389$. 

Table \ref{table-compare-sampling} shows the Monte Carlo estimates of one trained wave function (5 units, 2DEG with Rashba) using four different sampling schemes. The sampling schemes achieve the same energies within error bars, which validates the sampling implementation. 

\begin{table}[ht]
  \caption{Monte Carlo evaluation of energies. Different sampling schemes give same energies within error bars.}
  \label{table-compare-sampling}
  \centering
  \begin{tabular}{p{0.13\linewidth}llll}

    \toprule
  Sampling      & Energy (Ha) & Rashba (Ha) & Kinetic (Ha) & Potential (Ha) \\
    \midrule

   Continuous, Langevin & -1.63069(6) & -0.21122(3)  & 0.4548(1)   &  -1.8743(2) \\
           \midrule

      Discrete, Langevin & -1.63084(4)  & -0.21120(4)  & 0.4550(1)  & -1.8747(1) \\
           \midrule
        Continuous, Gaussian & -1.63080(5) & -0.21122(4)  & 0.4549(2)  & -1.8745(2) \\
  \midrule
    Discrete, Gaussian & -1.63084(4) & -0.21128(4)  & 0.4553(2)  & -1.8749(2) \\
    \bottomrule
  \end{tabular}
\end{table}

\clearpage
\section{Training details}

Table \ref{table-hyperparameters} are the hyperparameters used for both $\text{H}_3$ and 2DEG. Tables \ref{table-h3-hyperparameters} and \ref{table-2deg-hyperparameters} show hyperparameters used for $\text{H}_3$ and 2DEG, respectively. Spin sampling is off for collinear and projected collinear ansatzes. The energy MCMC evaluations were run for 50,000 and 100,000 steps, with statistics reported for the last 10,000 and 20,000 steps, for $\text{H}_3$ and 2DEG, respectively. The spin proposal width $w$ determines $\kappa$ in the VMF distribution as $\kappa = \frac{1}{2 w^2}$.

\begin{table}[ht]
  \caption{Default training hyperparameters.}
  \label{table-hyperparameters}
  \centering
  \begin{tabular}{clc}
    \toprule
       & Parameter  & Value  \\
    \midrule
         & Complex orbitals & True  \\
         & Layers & 2 \\
         & No. determinants & 1 \\
          & Clip local energy & 5.0 \\
      MCMC     & Proposal standard deviation & 1 \\
       MCMC     & Spin proposal width & 0.6 \\
       MCMC     & Adjust width & Off \\
        KFAC     & Momentum & 0 \\
         KFAC     & Covariance moving average decay & 0.95 \\
         KFAC     & Norm constraint & 1e-3 \\
         KFAC     & Damping & 1e-3 \\
                  LR     & Decay & 1 \\
         LR     & Delay & 10000 \\
    \bottomrule
  \end{tabular}
\end{table}

\begin{table}[ht]
  \caption{Training hyperparameters for $\text{H}_3$.}
  \label{table-h3-hyperparameters}
  \centering
  \begin{tabular}{clc}
    \toprule
       & Parameter  & Value  \\
    \midrule
         & Batch size & 32000  \\
         & x64 & True \\
          & Pretrain iterations & 1000 \\
          & Train iterations & 40000 \\
                      MCMC     & Steps & 10 \\
                  LR     & Rate & 0.05 \\

    \bottomrule
  \end{tabular}
\end{table}

\begin{table}[ht]
  \caption{Training hyperparameters for 2DEG.}
  \label{table-2deg-hyperparameters}
  \centering
  \begin{tabular}{clc}
    \toprule
       & Parameter  & Value  \\
    \midrule
         & Batch size & 2048 \\
         & x64 & False \\
          & Pretrain iterations & 0 \\
          & Train iterations & 90000 \\
                   MCMC     & Steps & 20 \\
                  LR     & Rate & 0.1 \\

    \bottomrule
  \end{tabular}
\end{table}

\clearpage

\subsection{Computational details}

We ran the $\text{H}_3$ experiments on one A100 GPU and the 2DEG experiments on one L40 GPU. Table \ref{table-compute-req} shows the time per training step for the experiments. We use projected spinor and spinor to compare the effect of spin sampling on computational requirement. For $\text{H}_3$ the spin sampling added around 0.006 s/training iteration, and for 2DEG the spin sampling added around 0.02 s/training iteration. Because we move the spin of one electron at a time, the spin sampling is implemented as $O(n m)$ where $n$ is the number of electrons and $m$ is the number of MCMC steps. The cost of spin sampling may be optimized in future work. Comparing the 2DEG with and without Rashba, the Rashba energy calculation adds about 0.033 s/training iteration. The Rashba energy calculation is order $n$ times the cost of a gradient calculation $\frac{\partial \Psi}{\partial R}$.  

\begin{table}[ht]
  \caption{Computational cost per training step.}
  \label{table-compute-req}
  \centering
  \begin{tabular}{lcccllc}
    \toprule
    System   & Nelec  & MCMC step & Batch size & x64 & MCMC step & Train time (s) \\
    \midrule
    $\text{H}_3$   &  3 & 10 & 32000 & True & $\text{P}_S$ spinor & 0.053 \\
   &   &  &  &  &Spinor &  0.059 \\
   2DEG   &  10 & 20 & 2048  & False & $\text{P}_S$ spinor &  0.057 \\
   &   &  &  &  & Spinor &  0.075 \\
   2DEG w/ Rashba   &  10  & 20 &  2048 & False & $\text{P}_S$ spinor &  0.09 \\
      &   &  &   & & Spinor &  0.11 \\
    \bottomrule
  \end{tabular}
\end{table}

\section{Additional experiments}

We compared our noncollinear implementation (called SpinorNet) with FermiNet for $\text{H}_3$ and with DeepSolid \cite{li-2022-ab-initio} for periodic H chain. Figs.~\ref{fig:h3-vs-ferminet} and \ref{fig:hchain-vs-deepsolid} show the results for $\text{H}_3$ and H chain, respectively. SpinorNet achieves energies on-par with FermiNet and DeepSolid, which verifies our network and is expected given the theoretical bounds of the ansatzes. We trained all networks for 30,000 iterations. We used 256 single units, 32 double units for SpinorNet and the default hyperparameters of FermiNet and DeepSolid. 

\begin{figure}[ht!]
\includegraphics{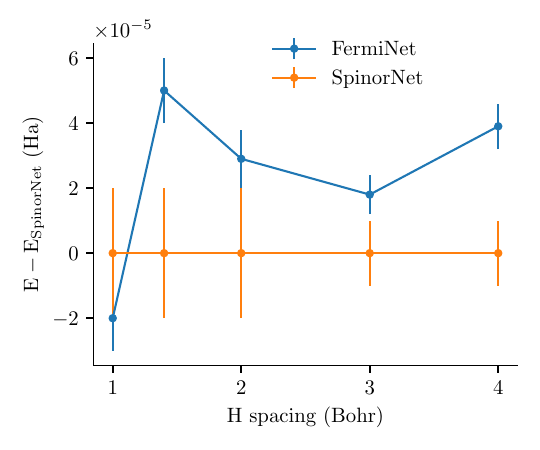}
\caption{SpinorNet energies are on-par with FermiNet-16dets for $\text{H}_3$ molecule.}
\label{fig:h3-vs-ferminet}
\end{figure}

\begin{figure}[ht!]
\includegraphics{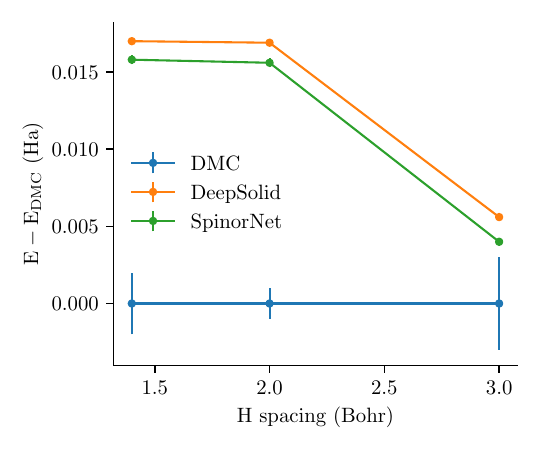}
\caption{SpinorNet energies are on-par or lower with DeepSolid for periodic H chain with 10 H atoms.}
\label{fig:hchain-vs-deepsolid}
\end{figure}

\clearpage

\endgroup  

\end{document}